# Predicting Electronic Structure Properties of Transition Metal Complexes with Neural Networks

Jon Paul Janet[1] and Heather J. Kulik[1,*]

[1]*Department of Chemical Engineering, Massachusetts Institute of Technology, Cambridge, MA, 02139*

ABSTRACT: High-throughput computational screening has emerged as a critical component of materials discovery. Direct density functional theory (DFT) simulation of inorganic materials and molecular transition metal complexes is often used to describe subtle trends in inorganic bonding and spin-state ordering, but these calculations are computationally costly and properties are sensitive to the exchange-correlation functional employed. To begin to overcome these challenges, we trained artificial neural networks (ANNs) to predict quantum-mechanically-derived properties, including spin-state ordering, sensitivity to Hartree-Fock exchange, and spin-state specific bond lengths in transition metal complexes. Our ANN is trained on a small set of inorganic-chemistry-appropriate empirical inputs that are both maximally transferable and do not require precise three-dimensional structural information for prediction. Using these descriptors, our ANN predicts spin-state splittings of single-site transition metal complexes (i.e., Cr-Ni) at arbitrary amounts of Hartree-Fock exchange to within 3 kcal/mol accuracy of DFT calculations. Our exchange-sensitivity ANN enables improved predictions on a diverse test set of experimentally-characterized transition metal complexes by extrapolation from semi-local DFT to hybrid DFT. The ANN also outperforms other machine learning models (i.e., support vector regression and kernel ridge regression), demonstrating particularly improved performance in transferability, as measured by prediction errors on the diverse test set. We establish the value of new uncertainty quantification tools to estimate ANN prediction uncertainty in computational chemistry, and we provide additional heuristics for identification of when a compound of interest is likely to be poorly predicted by the ANN. The ANNs developed in this work provide a strategy for screening of transition metal complexes both with direct ANN prediction and with improved structure generation for ANN validation with first-principles simulation.

# 1. Introduction

High-throughput computational screening has become a leading component of the workflow for identifying new molecules[1-2], catalysts[3], and materials[4]. First-principles simulation remains critical to many screening and discovery studies, but relatively high computational cost of direct simulation limits exploration of chemical space to a small fraction of feasible compounds[5-6]. In order to accelerate discovery, lower levels of theory, including machine-learning models, have emerged as alternate approaches for efficient evaluation of new candidate materials[7]. Artificial neural networks (ANNs) have recently found wide application in the computational chemistry community[8-10]. Machine learning approaches were initially appreciated for their flexibility to fit potential energy surfaces and thus force field models[10-16]. Broader applications have recently been explored, including in exchange-correlation functional development[8, 17], orbital free density functional theory[18-19], acceleration of dynamics[20-22], and molecular[1-2] or heterogeneous catalyst[23] and materials[24-27] discovery, to name a few.

Essential challenges for ANNs to replace direct calculation by first-principles methods include the appropriate determination of broadly applicable descriptors that enable the use of the ANN flexibly beyond molecules in the training set, e.g. for larger molecules or for those with diverse chemistry. Indeed, the most successful applications of ANNs at this time beyond proof-of-concept demonstration in replacement of direct first-principles simulation have been in the development of force fields for well-defined compositions, e.g. of water[28-29]. Within organic chemistry, structural descriptors such as a Coulomb matrix[30] or local descriptions of the chemical environment and bonding[31-32] have been useful to enable predictions of energetics as long as a relatively narrow range of compositions is considered (e.g., C, H, N, O compounds). These observations are consistent with previous successes in cheminformatics for evaluating molecular similarity[33], force field development[34], quantitative structure-activity relationships[35], and group additivity[36] theories. For transition metal complexes, few force fields have been established that can capture a full range of inorganic chemical bonding[37], and the spin-state- and coordination-environment-dependence of bonding[38] suggests that more careful development of descriptors is required to broadly predict properties of mid-row transition metal complexes. Similarly, descriptors that worked well for organic molecules have been demonstrated to not be suitable in inorganic crystalline materials[39]. We have previously observed[40] a strong relationship between sensitivity of electronic properties (e.g., spin-state splitting) and the direct ligand-atom and ligand field strength[41-42] in transition-metal complexes. Since ligands with the same direct metal-bonding atom can have substantially different ligand-field strengths (e.g., C for both weaker field $CH_3CN$ versus strong-field CO), whereas distant substitutions (e.g., tetraphenylporphyrin vs. base porphine) will have no effect, a transition-metal complex descriptor set that carefully balances metal-proximal and metal-distant descriptors is needed.

Within transition metal chemistry and correlated, inorganic materials, a second concern arises for the development of ANN predictions of first-principles properties. Although efficient correlated wavefunction theory methods (e.g., MP2) may be straightforwardly applied to small organic molecules, such methods are not appropriate for transition metal complexes where best practices remain an open question[43]. Although promising avenues for ANNs include the mapping of lower-level theory results, e.g. from semi-empirical theory[44], to a higher-level one, as has been demonstrated on atomization energies[45] and more recently reaction barriers[46], suitable levels of theory for extrapolation are less clear in transition metal chemistry.

There remains uncertainty about the amount of HF exchange to include in study of transition metal complexes, with recommendations ranging from no exchange, despite disproportionate delocalization errors in approximate DFT on transition metal complexes[41, 47-48], to alternately low[49-51] or high[52] amounts of exact exchange in a system-dependent manner. Indeed, there has been much interest recently in quantifying uncertainty with respect to functional choice in energetic predictions[53-55], including through evaluation of sensitivity of transition metal complex predictions with respect to inclusion of exact exchange[40, 52]. Spin-state splitting is particularly sensitive to HF exchange fraction, making it a representative quantity for which it is useful to both obtain a direct value and its sensitivity to varying the exchange fraction. Thus, a machine-learning model that predicts spin-state ordering across HF exchange values will be useful for translating literature predictions or providing sensitivity measures on computed data.

Overall, a demonstration of ANNs in inorganic chemistry, e.g. for efficient discovery of new spin-crossover complexes[56-57], for dye-sensitizers in solar cells[58], or for identification of reactivity of open-shell catalysts[59] via rapid evaluation of spin-state ordering should satisfy two criteria: i) contain flexible descriptors that balance metal-proximal and metal-distant features and ii) be able to predict spin-state ordering across exchange-correlation mixing. In this work, we make progress toward both of these aims, harnessing cheminformatics-inspired transition metal-complex structure generation tools[60] and established structure-functional sensitivity relationships in transition metal complexes[40, 52] to train ANNs for transition metal complex property prediction.

The outline of the rest of this work is as follows. In Sec. 2 (Methods), we review the computational details of data set generation, we discuss our variable selection procedure, and we review details of the artificial neural network trained. In Sec. 3, we provide the Results and Discussion on the trained neural networks for spin-state ordering, spin-state exchange sensitivity, and bond-length prediction on both training-set-representative complexes and diverse experimental complexes. Finally, in Sec. 4, we provide our conclusions.

## 2. Methods

### 2.1 Test Set Construction and Simulation Details

*Data set construction*. Our training set consists of octahedral complexes of first-row transition metals in common oxidation states: $Cr^{2+/3+}$, $Mn^{2+/3+}$, $Fe^{2+/3+}$, $Co^{2+/3+}$, and $Ni^{2+}$. High-spin (H) and low-spin (L) multiplicities were selected for each metal from the ground, high-spin state of the isolated atom and the higher-energy, lowest-spin state within 5 eV that had a consistent *d*-orbital occupation for both states, as obtained from the National Institute of Standards and Technology atomic spectra database[61]. The selected H-L states were: triplet-singlet for $Ni^{2+}$, quartet-doublet for $Co^{2+}$ and $Cr^{3+}$, quintet-singlet for $Fe^{2+}$ and $Co^{3+}$, quintet-triplet for $Cr^{2+}$ and $Mn^{3+}$ (due to the fact that there is no data available for $Mn^{3+}$ singlets[61]), and sextet-doublet for $Mn^{2+}$ and $Fe^{3+}$.

A set of common ligands in inorganic chemistry was chosen for variability in denticity, rigidity, and size (nine monodentate, six bidentate, and one tetradentate in Figure 1 and Supporting Information Table S1). These ligands span the spectrochemical series from weak-field chloride (**1,** Cl⁻) to strong-field carbonyl (**6**, CO) along with representative intermediate-field ligands and connecting atoms, including S (**2**, SCN⁻), N (e.g., **9**, $NH_3$), and O (e.g., **14**, acac). All possible homoleptic structures with all metals/oxidation states were generated from ten of these ligands (90 molecules) using the molSimplify toolkit[60] (Supporting Information Table S2). Additional heteroleptic complexes (114 molecules) were generated with molSimplify with one mono- or bidentate axial ligand type ($L_{ax}$) and an equatorial ligand type ($L_{eq}$) of compatible denticity (ligands shown in Figure 1, schematic shown in Figure 2, geometries provided in the Supporting Information). We also selected 35 molecules from the Cambridge Structural Database[62] (Supporting Information Table S3).

*First-principles geometry optimizations*. DFT gas-phase geometry optimizations were carried out using TeraChem.[63-64] DFT calculations employ the B3LYP hybrid functional[65-67] with 20% Hartree-Fock (HF) exchange ($a_{HF}$ = 0.20) and a variant[40] ($a_{HF}$ = 0.00 to 0.30 in 0.05 increments) that holds the semi-local DFT portion of exchange in a constant ratio. We calculate and predict sensitivities with respect to HF exchange, $\frac{\partial \Delta E_{H-L}}{\partial a_{HF}}$, as approximated from linear fits, in units of kcal/mol·HFX$^{-1}$, where 1 HFX corresponds to varying from 0% to 100% HF exchange. B3LYP[65-67] is chosen here due to its widespread use and our prior experience[40] with tuning it to study HF exchange sensitivity, where we also observed[40] similar behavior with other GGA hybrids, e.g. PBE0, as long as the same HF exchange fraction was compared.

The composite basis set used consists of the LANL2DZ effective core potential[68] for transition metals and the 6-31G* basis for the remaining atoms. All calculations are spin-unrestricted with virtual and open-shell orbitals level-shifted[69] by 1.0 and 0.1 eV, respectively, to aid self-consistent field (SCF) convergence to an unrestricted solution.

For all training and test case geometry optimizations, default tolerances of 10$^{-6}$ hartree for SCF energy changes between steps and a maximum gradient of 4.5x10$^{-4}$ hartree/bohr were employed, as implemented in the DL-FIND interface[70] with TeraChem. Entropic and solvent effects that enable comparison to experimental spin-state splittings have been omitted, and we instead evaluate the DFT adiabatic electronic spin state splitting, as in previous work because our goal is to predict DFT properties and sensitivity to functional choice.[40, 71] In high-throughput screening efforts ongoing in our lab, entropic and solvent effects that influence catalytic and redox properties will be considered in the future.

For each molecular structure (90 homoleptic, 114 heteroleptic) 14 geometry optimizations were carried out at 7 exchange fractions (from 0.00 to 0.30) and in high- or low- spin, for a theoretical maximum of 2856 geometry optimizations. In practice, 166 structures were excluded due to i) large spin contamination, as defined by an expectation value of $<\hat{S}^2>$ that deviated more than 1 $\mu_B$ from the exact value (< 1%, 26 of 2856, see Supporting Information Table S4), ii) dissociation in one or both spin states, especially of negatively charged ligands, leading to loss of octahedral coordination (4%, 126 of 2856, see Supporting Information Table S5), or iii) challenges associated with obtaining a stable minimized geometry (< 1%, 14 of 2856, see Supporting Information Table S2). Eliminating these cases produced a final data set of 2,690 geometry optimizations (structures and energetics provided in Supporting Information, as outlined in Supporting Information Text S1). Although these excluded cases are a fraction of our original data set, they highlight considerations for application of the ANN in high-throughput screening: highly negatively charged complexes should be avoided, and single point DFT calculations should be used to confirm that a high-fitness complex does not suffer from large $<\hat{S}^2>$ deviations.

**2.2. Descriptor Selection**

High-throughput screening of transition-metal complex properties with direct prediction from an ANN requires mapping of an empirical feature space that represents the complex, $\mathcal{X}$, to quantum-mechanical predictions. This feature space should be balanced to avoid i) too few descriptors with insufficient predictive capability or ii) too many descriptors that lead to over-fitting of the ANN. Molecular descriptors[72] that have been used for parameterizing chemical space include: atomic composition, electronegativity[32], formal charges, and representations of the geometric structure. This last class of descriptors may be divided into those that depend either on 3D structural information[13, 18, 73-75] or on graph-theoretic connectivity maps[76] (e.g., the Randić[77], Wiener shape[78], or Kier[79] indices). Graph-theoretic methods are preferable to 3D structural information to avoid sensitivity to translation/rotation or molecule size[80], though we note that subsystem descriptors[13, 75, 81] and element-specific pairwise potentials[74, 80] have been employed successfully to overcome some challenges. A secondary reason to avoid use of 3D structural information is the implicit requirement of equilibrium geometries obtained from a geometry optimization, which are readily achieved with semi-empirical methods on small organic molecules[76] but would be prohibitive and error-prone for transition metal complexes.

We use L1-regularized, least absolute shrinkage and selection operator (LASSO) linear least-squares regression[82], as implemented in the glmnet[83] package in R3.2.5[84], to evaluate candidate descriptor sets. LASSO is used to reduce over-fitting, force the coefficients of the least-powerful indicators to zero, and avoid monotonic decrease of model error as feature space dimensionality increases. Given observed input-output pairs ($\mathbf{x}_i$, $\mathbf{y}_i$) for $i=1,\ldots,n$ with $\mathbf{x} \in \mathcal{X} \subset \mathbb{R}_i^m$ and $\lambda \in \mathbb{R}$, the output is modeled as:

$$\tilde{\mathbf{y}} = \beta^T \mathbf{X} + \beta_0 \mathbf{1} \qquad (1)$$

for $\beta, \beta_0 \in \mathbb{R}^m \times \mathbb{R}$, where:

$$\{\beta, \beta_0\} = \arg\min_{\{\beta, \beta_0\}} \left( \left\| \mathbf{y} - \beta^T \mathbf{X} - \beta_0 \mathbf{1} \right\|_2^2 + \lambda \left( \sum_{i=1}^{m} |\beta_i| \right) \right) \qquad (2)$$

The parameter λ is selected by ten-fold cross-validation with values typically between $10^{-1}$ and $10^{-6}$. Our descriptors include both continuous variables that are normalized and discrete variables that are described by zero-one binary coding (Supporting Information Table S6). Metal identity represents a descriptor best described by a set of discrete variables: 4 binary variables are chosen to represent Cr, Mn, Fe, and Ni, and Co corresponds to the case where all 4 variables are zero. This leads to a higher number of overall variables than for continuous descriptors (see Table 1).

Based on previous observations[40, 42], we hypothesize that spin-state ordering is predominantly determined by the immediate chemical environment around the metal center, potentially enabling predictive descriptors that are widely transferable across a range of molecule sizes. We compare 7 descriptor sets on the data and select the subset of descriptors that give the best simultaneous predictive performance for spin-state splitting, $\Delta E_{\text{H-L}}$, and its sensitivity with respect to HF exchange variation, $\frac{\partial \Delta E_{\text{H-L}}}{\partial a_{\text{HF}}}$, as indicated by the prediction root mean squared error (RMSE):

$$\text{RMSE} = \sqrt{\frac{1}{N}\sum_{i=1}^{N}(y_{i,\text{pred.}} - y_{i,\text{actual}})^2} \qquad (3)$$

When two variable sets perform comparably, we select the variable set that will enable broader application of the ANN. All sets include the metal identity as a discrete variable and metal oxidation state, ligand formal charge, and ligand denticity as continuous variables (Figure 3, some descriptors shown in Fig. 2). Set **a** represents our most specific model, where we explicitly code the full axial or equatorial ligand identity as a discrete variable, limiting the application of the model but producing one of the lowest RMSEs for $\Delta E_{\text{H-L}}$ and $\frac{\partial \Delta E_{\text{H-L}}}{\partial a_{\text{HF}}}$ (Table 1). Elimination of ligand identity in favor of ligand connecting atom elemental identity and total number of atoms in set **b** increases $\Delta E_{\text{H-L}}$ MSE slightly and decreases $\frac{\partial \Delta E_{\text{H-L}}}{\partial a_{\text{HF}}}$ MSE (see Table 1).

The shift from set **a** to **b** increases the model applicability but at the cost of omitting subtler ligand effects. For instance, ethylenediamine (**11**, en) and phenanthroline (**10**, phen) have the same ligand charge/denticity and direct ligand atom (N), making them equivalent in set **b** except for the larger size of phen. System size alone is not expected to be a good predictor of field strength (e.g., the small CO is one of the strongest field ligands). In set **c**, we introduce properties that depend on the empirical pairwise Pauling electronegativity difference ($\Delta \chi$) between the ligand connecting atom (LC) and any $i$th atom connected (CA) to it:

$$\Delta\chi_{\text{LC},i} = \chi_{\text{LC}} - \chi_i \tag{4}$$

These whole-complex differences include the maximum, max($\Delta\chi$), and minimum, min($\Delta\chi$), as well as sum:

$$\text{sum}(\Delta\chi) = \sum_{j\in\text{lig.}} \sum_{i\in\text{CA}} \Delta\chi_{\text{LC},i}^{j} \tag{5}$$

which is taken over the direct ligand atom and all atoms bonded to it for all ligands (lig.) in the complex. These additional set **c** descriptors reduce $\Delta E_{\text{H-L}}$ MSE slightly and decrease the $\dfrac{\partial \Delta E_{\text{H-L}}}{\partial a_{\text{HF}}}$ MSE to its lowest value (see Table 1). In set **d**, we eliminate min($\Delta\chi$), expecting it to be redundant with the max and sum, at the cost of a small increase in both MSEs.

Finally, in sets **e**-**g**, we replace ligand size (i.e., number of atoms) with general descriptors to enable prediction on molecules larger than those in any training set. For example, tetraphenylporphyrin will have comparable electronic properties to unfunctionalized porphyrin (**12**), despite a substantial size increase. In set **e**, we introduce the maximum bond order of the ligand connecting atom to any of its nearest neighbors, a measure of the rigidity of the ligand environment, which is zero if the ligand is atomic (see Supporting Information Table S1). In set **f**, we eliminate the number of atoms and bond order metric, replacing them with a broader measure of the ligand geometry adjacent to the metal. After trial and error, we have selected the truncated Kier shape index[79], $^2\kappa$, which is defined by the inverse ratio of the square of number unique paths of length two ($^2P$) in the molecular graph of heavy atoms to the theoretical maximum and minimum for a linear alkane with the same number of atoms:

$$^2\kappa = \frac{2\,^2P_{\max}\,^2P_{\min}}{\left(^2P\right)^2} \tag{6}$$

and set to zero for any molecules that do not have paths of length two. The truncation means that only the ligand atoms within three bonds of the connecting atom are included in the graph. The set **f** MSEs are comparable to or a slight increase from sets with molecule size, but they beneficially eliminate system size dependence. In set **g**, we reintroduce the bond order metric as well, providing the lowest MSEs except for set **a** or **c**, both of which are much less transferable than set **g**. Thus, the comparable performance of set **g** to a full ligand descriptor (set **a**) supports our hypothesis that a combination of metal-centric and ligand-centric in a heuristic descriptor set can be predictive and transferable. This final feature space is 15-dimensional with five per-complex descriptors and five per-ligand descriptors for each equatorial or axial ligand (see Table 2 for ranges of values and descriptions). A comparison of all errors and weights of variables across the seven data sets is provided in Supporting Information Tables S7-S20 and Figure S1.

**2.3 Training and Uncertainty Quantification of ML models**

ANNs enable complex mapping of inputs to outputs[85] beyond multiple linear regression and support the use of both discrete (i.e., binary choices such as metal identity) and continuous (e.g., the % of HF exchange) variables. Here, we apply an ANN with an input layer, two intermediate hidden layers, and an output layer (Figure 2). The network topology was determined by trial and error, with additional hidden layers yielding no improved performance. All analysis is conducted in R 3.2.5[84], using the h2o[86] package with tanh non-linearity and linear output. Network weights and full training and test data are provided in the Supporting Information.

As with many ML models, ANNs are sensitive to over-fitting due to the number of weights to be trained[87]. We address overfitting using dropout[88-89], wherein robustness of the fit is improved by zeroing out nodes in the network with an equal probability, $p_{drop}$, at each stage of training (5% for spin-state splitting, 15% for HF exchange sensitivity, and 30% for bond lengths, selected by trial and error). Dropout has been shown to address overfitting when training feedforward ANNs on small datasets[89], with larger values of $p_{drop}$ giving more aggressive regularization that worsens training errors but improves test errors. We use L2 weight regularization with a fixed penalty weight $\lambda$, as is applied in standard ridge regression, with an effective loss function for training:

$$\{\mathbf{W}\} = \arg\min_{\{\mathbf{W}\}} \left( \sum_{i=1}^{N} \left( \mathbf{y}_n - \tilde{\mathbf{y}}(\mathbf{x}_n) \right)^2 + \lambda \sum_{l=1}^{L} \left( \|\mathbf{W}_l\|_2^2 + \|\mathbf{b}_l\|_2^2 \right) \right) \quad (7)$$

Here, $\mathbf{W}_l$ refers to the weights from layer $l$ to $l+1$, $\mathbf{b}_l$ are biases at layer $l$, $\tilde{\mathbf{y}}(\mathbf{x}_n)$ is the ANN prediction for the input-output pair $(\mathbf{x}_n, \mathbf{y}_n)$, and the sums run over $N$ training pairs and $L$ layers.

During network training, we randomize the order of data points and partition the first 60% as training data and the last 40% for testing. Dropout networks, consisting of two hidden layers of 50 nodes each, are trained on the data set for varying values of $\lambda$ ranging from $10^{-1}$ to $10^{-6}$ using 10-fold cross validation. For each $\lambda$, the training data is partitioned into ten groups, a network is trained on nine of the groups and scored based on eq. 7 on the left-out group to select the best regularization parameter: $5\times10^{-4}$ for spin-state splitting, $10^{-2}$ for HF exchange sensitivity, and $10^{-3}$ for bond lengths. We varied and optimized[90] the learning rate between 0.05 and 1.5, and optimal rates were selected as 1.0 (bond lengths) and 1.5 (spin-state splitting or HF exchange sensitivity). We use batch optimization for training (batch size = 20) for 2000 epochs. The training algorithm minimizes eq. 7 over the training data using stochastic gradient descent[90-93].

It has not been possible to estimate ANN model uncertainty[88, 94] with the possible exception of bootstrapping[95] by training the ANN on numerous subsamples of available training data.

Model uncertainty will be due to either high-sensitivity to descriptor changes or test molecule distance in chemical space to training data (see also Sec. 3). Recent work[87] showed that minimization of the loss function in eqn. 7 is equivalent to approximate variational optimization of a Gaussian process (GP), making previously suggested ANN sampling for different dropout realizations[88] a rigorously justified[87] model uncertainty estimate.

We sample $J$ distinct networks (in this work, $J$=100) with different nodes dropped at the optimized weights and average over the predictions:

$$\tilde{\mathbf{y}}(\mathbf{x}_n) = \frac{1}{J}\sum_{j=1}^{J}\tilde{\mathbf{y}}_j(\mathbf{x}_n) \quad (8)$$

The ANN predictive variance is[87]:

$$\text{var}(\tilde{\mathbf{y}}(\mathbf{x}_n)) \approx \tau^{-1}\mathbf{I} + \frac{1}{J}\sum_{j=1}^{J}\left(\mathbf{y}_j(\mathbf{x}_n)^T \mathbf{y}_j(\mathbf{x}_n) - \tilde{\mathbf{y}}(\mathbf{x}_n)^T \tilde{\mathbf{y}}(\mathbf{x}_n)\right) \quad (9)$$

Here, $\tau$ is

$$\tau = \frac{(1-p_{drop})l^2}{2N\lambda} \quad (10)$$

where $N$ is the number of training data points, and $l$ is a model hyperparameter for the GP that affects the estimation of predictive variance but does not enter into the ANN training. The contribution of $\tau$ in eqn. 9 is a baseline variance inherent in the data, whereas the second term represents the variability of the GP itself. We obtain $\tau$ values of 0.6 for spin-state splitting, 0.07 for HF exchange sensitivity, and 10000 for bond lengths (see Sec. 3). We choose $l$ by maximizing the log predictive likelihood of the corresponding GP based on the training data (details are provided in the Supporting Information Text S2).

We selected an ANN based on the successful demonstrations[11, 14, 96] of ANN-based models for predicting quantum chemical properties but also provide a comparison to two other common machine learning models[82]: kernel ridge regression (KRR) and a support vector regression model (SVR), both using a square-exponential kernel. We used the R package kernlab[97] and selected hyperparameters (the width of the kernel, and the magnitude of the regularization parameters which are given in the Supporting Information Table S21) using a grid search and ten-fold cross-validation using the R package CVST[98]. We also compared training on our descriptor set to a KRR model with a kernel based on the L1 distance between sorted Coulomb matrix representations[80], as demonstrated previously[45, 96].

## 3. Results and Discussion

### 3.1 Overview of Data Set Spin-State Energetics

Analysis of the qualitative and quantitative features of the spin-state splitting data set motivates the training of an ANN to move beyond ligand field arguments. We visualize qualitative ground states (i.e., high-spin or low-spin) for the homoleptic subset of the data using a recursive binary tree (Figure 4, descriptor definitions provided in Table 2), as previously outlined[99] and implemented in the open source rpart package[100] for R 3.2.5[84]. A recursive binary tree is a list of "branches" of the data ordered by statistical significance that gives the most homogeneous final "leaves" (here, with at least 10 data points) after a given number of permitted divisions (here, 6). Using descriptor set **g**, the data are partitioned into branches by testing which descriptors provide the "best" division to produce majority high- or low-spin states in leaves based on the concept of information impurity[100] and pruning to remove statistically insignificant branches. The resulting electronic structure spectrochemical "tree" simultaneously addresses metal-specific strengths of ligands and exchange-correlation sensitivity. As expected, strong field direct carbon ligands (no Cl, N, O or S in Figure 4) provide the root division of the tree, producing low-spin ground states for 92% of all Mn, Fe, and Co complexes (far right box on the third tier in Figure 4). Next level divisions include the M(II) oxidation state for $a_{HF} > 0.05$ that are predominantly (96%) high-spin. Spin-state ordering is well-known[40, 52] to be sensitive to HF exchange, and the tree reveals $Mn^{3+}$ with nitrogen ligands to have the strongest $a_{HF}$ dependence, since they are 69% high-spin for $a_{HF} > 0.1$ but 90% low-spin for $a_{HF} \leq 0.1$. Extension of the recursive binary tree to heteroleptic compounds produces a second-level division based on sum($\Delta\chi$), validating the relevance of the identified electronegativity descriptors for predicting heteroleptic spin-state ordering (Supporting Information Figure S2).

Quantitatively, the maximum $\Delta E_{H-L}$ in the data set is 90.7 kcal/mol for the strong-field Co(III)(misc)$_6$ complex at $a_{HF} = 0.00$, and the minimum value is -54.2 kcal/mol for the weak-field Mn(II)(NCS$^-$)$_6$ at $a_{HF} = 0.30$. These extrema are consistent with i) the ordering of metals in the spectrochemical series[38] and ii) the uniform effect of stabilizing high-spin states with increasing HF exchange. By comparing compound trends in the data set, we are able to identify whether additivity in ligand field effects, which has been leveraged previously in heuristic DFT correction models[101-103], is a universally good assumption. For the Fe(III)(Cl$^-$)$_{6-n}$(pisc)$_n$ complexes (denoted **1-1** through **3-3** in Figure 5), increasing $n$ from 0 to 2 through the addition of two axial pisc ligands increases the spin-state splitting by 15.1 kcal/mol per replaced chloride. Transitioning to a complex with all equatorial pisc ligands ($n$=4) increases the spin-state splitting by only 10.4 kcal/mol per additional ligand, and the homoleptic structure pisc ($n$=6) only adds 7.5 kcal/mol per additional ligand beyond the $n$=4 case. An additive model cannot precisely reproduce diminishing ligand effects. As a stronger example for the need for nonlinear models such as an ANN, replacing two axial ligands from the strong-field Mn(II)(CO)$_6$ complex with the weaker-field NCS$^-$ (**6-6** and **6-7** in Supporting Information Figure S3) alters $\Delta E_{H-L}$ by < 1 kcal/mol, as strong-field ligands (e.g., CO, CN$^-$) have an overriding effect on spin-state splitting.

**3.2 Spin-State Splittings from an ANN**

Motivated by non-linear effects in ligand additivity, we trained an ANN using a heuristic descriptor set (see Sec. 2.2) to predict qualitative spin-state and quantitative spin-state splitting. The ANN predicts the correct ground state in 98% of the test cases (528 of 538) and 96% of training cases (777 of 807). All of the misclassifications are for cases in which DFT $\Delta E_{H-L}$ is < ± 5 kcal/mol (Supporting Information Table S22). The ANN spin-state prediction errors are not sensitive to HF exchange mixing, and thus our trained ANN is able to predict ground states of transition metal complexes from the pure GGA limit to hybrids with moderate exchange.

We assess quantitative performance with root mean squared errors (RMSE) of the ANN (eqn. 3), overall and by metal (Figure 6, Supporting Information Table S21, and Supporting Information Figures S3-S6). The comparable RMSE of 3.0 and 3.1 kcal/mol for the test and training data, respectively, indicate an appropriate degree of regularization. The ANN predicts DFT spin-state splittings within 1 kcal/mol (i.e., "chemical accuracy") for 31% (168 of 538) of the test data and within 3 kcal/mol (i.e., "transition metal chemical accuracy[104]" for 72% (389 of 538) of the test data. Only a small subset of 49 (4) test molecules have errors above 5 (10) kcal/mol, and correspond to strong-field Co and Cr complexes, e.g., $Cr(II)(NCS^-)_2(pisc)_4$ (Supporting Information Figure S5). The model is equivalently predictive for homoleptic and heteroleptic compounds at 2.2 and 2.3 kcal/mol average unsigned error respectively.

The training and test RMSEs broken down by metal reveal comparable performance across the periodic table (Figure 6). Slightly higher test RMSEs (maximum unsigned errors) for Co and Fe complexes at 3.8 (15.7) and 3.3 (13.0) kcal/mol, respectively, are due to the train/test partition and more variable ligand dependence of spin-state ordering in these complexes (Figure 6 and Supporting Information Table S23). When the ANN performs poorly, the errors are due to both under- and over-estimation of $\Delta E_{H-L}$ for both strong- and weak-field ligands, regardless of HF exchange fraction: e.g., $\Delta E_{H-L}$ for Co(III)(CN$^-$)$_6$ at $a_{HF}$ = 0.00 and Co(III)(en)$_3$ at $a_{HF}$ = 0.20 are overestimated by 14 and 9 kcal/mol, respectively, but $\Delta E_{H-L}$ for Fe(III)(Cl$^-$)$_6$ at $a_{HF}$=0.10 and Co(II)(H$_2$O)$_2$(CN$^-$)$_4$ at $a_{HF}$ = 0.30 and are underestimated by 9 and 7 kcal/mol, respectively.

Quantified uncertainty estimates correspond to a baseline standard deviation in the model of approximately 1.5 kcal/mol ($\sqrt{\tau^{-1}}$) and a mean total estimated standard deviation across the training and test cases of 3.8 and 3.9 kcal/mol, respectively (see sec 2.3 and error bars on Figure 5). These credible intervals are not rigorously confidence intervals but can highlight when prediction uncertainty is high: a ±1 (±2) standard deviation (std. dev.) interval on ANN predictions captures 83% (98%) of computed values for test set (see Supporting Information Figure S7). Highest std. dev. values of around 5 kcal/mol are observed for Fe(II) and Mn(II) complexes and the lowest are around 3 for Cr and Co complexes (see Supporting Information). A single std. dev. around the ANN prediction contains the calculated $\Delta E_{H-L}$ for 26 of 29 Fe(III) complexes at $a_{HF}$=0.20 but misses heteroleptic oxygen coordinating complexes, **13-13** and **14-1**, and underestimates the effect of C/N ligands in **3-7** (Figure 5). The model performs consistently across different ligand sizes, from porphyrin Fe(III) complexes (**12-13**, **12-5**) to Fe(II)(NH$_3$)$_6$ and Fe(II)(CO)$_6$ (**6-6** and **8-8**). For ligand-specific effects, the ANN performs well, reversing splitting magnitude as equatorial and axial ligands are swapped (e.g., **1-3** versus **3-1**).

Review of other metals/oxidation states reveals comparable performance for cases where the high-spin state is always favored (e.g., Mn(II), Cr(III), or Ni(II)), low-spin state is always favored (e.g., Cr(III)), and those where ligands have strong influence over the favored spin state (e.g., Fe(II) and Cr(II)) (see Supporting Information Figure S3-S6). For instance, metal-specific effects examined through comparison of M(II)(CO)$_6$ complexes (Figure 7) reveal good ANN performance both for where the strong-field ligand strongly favors the low-spin state (i.e., Fe and Ni) and where the spin-states are nearly degenerate (i.e., Cr, Mn, Co). The trends outlined here for 20% HF exchange hold at other exchange mixing values (Supporting Information Table S22). Thus, our ANN trained on a modest data set with heuristic descriptors predicts spin-state splitting within a few kcal/mol of the DFT result.

Comparing our results to KRR, SVR, and LASSO regression reinforces the choice of an ANN (Table 3 and Supporting Information Figure S8). The ANN outperforms KRR with either our descriptor set or the sorted Coulomb matrix descriptor both on the full data set or at fixed HF exchange (Supporting Information Text S3). The ANN also performs slightly better than SVR on test data with our descriptors. Linear LASSO regression was employed for feature selection (Sec. 2.2) but is outperformed by all other methods (Table 3). We will revisit the performance of these models on a more diverse molecule test set in Sec. 3.5 to assess the question of transferability.

**3.3 Predicting Exchange Sensitivity with an ANN**

Spin-state splittings exhibit high sensitivity to exchange[40, 52] with linear behavior that we previously identified[40] to be strongly dependent on direct ligand identity and field strength when we compared a set of Fe complexes. Over this data set, computed exchange sensitivities are indeed linear, ranging from -174 kcal/mol·HFX$^{-1}$ for strong-field Fe(II)(CO)$_6$ to -13 kcal/mol·HFX$^{-1}$ for weak-field Cr(III)(en)$_2$(NH$_3$)$_2$. Cr(III) is the least exchange-sensitive metal in our test set, whereas Fe(II) and Mn(II) are the most sensitive (Supporting Information Table S24 and Figure S9).

We therefore generalize previous observations[40] in an ANN that predicts HF exchange sensitivity of spin-state ordering, $\frac{\partial \Delta E_{H-L}}{\partial a_{HF}}$, using the same descriptors as for direct spin-state splitting, excluding only $a_{HF}$. The smaller size of this data set (1/7 the size of the $\Delta E_{H-L}$ data set) leads to overfitting, with lower RMSE values of 13 kcal/mol·HFX$^{-1}$ for the training data versus 22 kcal/mol·HFX$^{-1}$ for the test set (Table 4, Supporting Information Figure S10 and Table S25). Although results are reported in units of HFX (from 0 to 100% exchange), for typical 20% variation in exchange, a 20 kcal/mol·HFX$^{-1}$ sensitivity error only corresponds to a 4 kcal/mol energy difference. Both maximum unsigned errors (UE) and RMSEs are largest for Mn(II/III) and Cr(II) complexes, with the largest case producing an 92 kcal/mol·HFX$^{-1}$ underprediction for Mn(III)(H$_2$O)$_4$(pisc)$_2$. Overall, the ANN prediction errors are less than less than 20 (40) kcal/mol·HFX$^{-1}$ for 65% (95%) of the test data. The ANN provides a valuable strategy for predicting exchange sensitivity, reproducing nonmonotonic and nonconvex ligand sensitivity in heteroleptic compounds: a Fe(III) complex with ox, **16**, and NCS$^-$, **7**, ligands is more sensitive to HFX than the respective homoleptic complexes (Figure 8, other metals in Supporting Information Figures S11-S14).

Uncertainty intervals of ANN predictions for HFX sensitivity yield a narrow range from 14 kcal/mol·HFX$^{-1}$ to 17 kcal/mol·HFX$^{-1}$. For the 29 Fe(III) complexes studied, 23 (80%) of the ANN credible intervals span the computed exchange sensitivity (Figure 8). Across the full metal and oxidation state data set, 70% (83%) of the computed data is contained by ± 1 (± 2) std. dev. intervals (Figure 8 and Supporting Information Figure S15). This performance can be further improved by extending the training data. Exchange-sensitivity provides value both for extrapolation of computed (see Sec. 3.6) or literature values obtained at an arbitrary exchange mixing and in identification of cases of high-sensitivity to DFT functional choice.

**3.4 Predicting Equilibrium Geometries with an ANN**

Using our descriptor set, we trained an ANN on the minimum metal-ligand bond distances for both low-spin and high-spin geometries (min($R_{LS/HS}$)), which only differ from the exact metal-ligand bond length for distorted or heteroleptic compounds. This ANN for bond length prediction extends capabilities we have recently introduced for generating high-quality transition metal complex geometries[60] in order to enable spin-state dependent predictions without requiring extended geometry-optimization. Furthermore, comparison of adiabatic and vertical spin-state splittings computed either at the low- or high-spin optimized geometries reveals that the vertical splitting at the HS geometry is indistinguishable from the adiabatic splitting, but the LS geometry vertical splitting favors the LS state by 10-30 kcal/mol, increasing with $a_{HF}$ (Figure 9). Thus, if the ANN bond length predictions are accurate, adiabatic spin-state splittings can be obtained from DFT single points at ANN-predicted HS-only or both LS/HS geometries.

Metal-ligand bond distances in the $a_{HF}$=0.20 data set vary from min($R_{LS}$)=1.81 Å (in Fe(II)(pisc)$_2$(Cl$^-$)$_4$) to min($R_{HS}$)=2.55 Å (in Fe(III)(Cl$^-$)$_6$). The metal-ligand bond length ANN produces comparable RMSE across training (0.02 Å for HS and LS) and test (0.02 Å for LS and 0.03 Å for HS) data with comparable errors regardless of metal identity and oxidation- or spin-state (Supporting Information Table S26-28 and Figures S16-27). ANN bond length std. devs. Range from 0.026 to 0.045 Å with a ~0.01 Å baseline contribution. For low-spin (high-spin) complexes, 79% (81%) and 96% (96%) of the calculated values fall within one and two std. dev. of ANN-predicted bond lengths, respectively (Supporting Information Figures S20 and S26).

The ANN overestimates bond lengths of low-spin Fe(III) complexes by more than a full standard deviation for seven cases, e.g., underestimating Fe-C distances in CN (**7-5, 13-5**) and pisc (**3-7, 3-13**) complexes (Figure 10). However, it also reproduces subtle trends, e.g. replacing axial ligands in homoleptic LS Fe(III)(pisc)$_6$ (**3-3** in Figure 10, min($R_{LS}$)=1.92 Å) with Cl$^-$ increases the minimum bond distance to 1.94 Å (**3-1** in Figure 10), but replacing equatorial pisc ligands instead with Cl⁻ (**1-3** in Figure 10) decreases the minimum bond distance to 1.90 Å, a feature reproduced by the ANN. Non-additive bond length effects motivate the use of the ANN in initial geometry construction[60]. Indeed, when we use ANN-predicted metal-ligand bond lengths in structure generation instead of our previous strategy based on a discrete database of DFT bond lengths[60], we reduce the metal-ligand component of the gradient by 54-90% (Supporting Information Text S4, Figure S28 and Table S29). The ANN-predicted bond lengths and spin states are now available in molSimplify[60] as an improved tool for structure generation.

**3.5 Expanding the Test Set with Experimental Transition Metal Complexes**

In order to test the broad applicability of the trained ANNs, we selected 35 homoleptic and heteroleptic octahedral complexes from the Cambridge Structural Database[62] (CSD) with a range of metals (Cr to Ni) and direct ligand atom types (N, C, O, S, Cl) (Supporting Information Table S30). The CSD test cases span a broader range of compounds than the training set, containing i) larger macrocycles, e.g. substituted porphyrins (tests 9, 25), clathrochelates (test 16), phthalocyanines (tests 4, 7), and cyclams (tests 5, 12, 14, 17, 24, 29, and 33, 12 and 33 shown in Figure 11) and ii) coordination combinations or functional groups, e.g., OCN in test 30, absent from the training set. Indeed, large CSD test molecule sizes, e.g. up to 103 atoms in a single equatorial ligand, further motivates our relatively size-independent descriptor set over forms that do not scale well with molecule size.

The ANN predicts CSD test case spin-state splittings within 5 kcal/mol for 15 of the 35 complexes, an overall mean unsigned error of 10 kcal/mol, and RMSE of 13 kcal/mol (See Supporting Information Table S31). The large RMSE is due in part to poor performance on early-transition-metal cyclams (red symbols in left panel of Figure 12) for which the ANN overestimates spin-state splitting by about 30 kcal/mol (Cr-cyclams, tests 12 and 33 in Figure 11). The ANN predicts spin-state splittings within around 3 kcal/mol for several non-macrocyclic complexes that are better represented in the training data (e.g., test cases 8 and 31 in Figure 11). The correct ground state is assigned in 90% of CSD test cases (96% after excluding cyclams); the only incorrect, non-cyclam spin state assignment is a spin-crossover complex, test 25 (calculated $\Delta E_{H-L}$ = -0.2 kcal/mol). Compared to other machine learning models (KRR and SVR), the ANN is more transferable to dissimilar CSD structures (Table 3), outperforming the next-best model, SVR, by 30%. The relative success of the ANN on the CSD data is partially attributable to the use of dropout regularization, which has been shown[89] to improve robustness.

The observation of good performance with reasonable similarity between CSD structures and the training data but poor performance when the CSD structure is not well-represented motivates a quantitative estimate of test compound similarity to training data. We first computed overall molecular similarity metrics (e.g., FP2 fingerprint via Tanimoto[33, 105], as implemented in OpenBabel[106-107]) but found limited correlation ($R^2$=0.1) to prediction error (see Supporting Information Figure S29 and Text S5). Comparing the Euclidean and uncentered Pearson distances in descriptor space between the CSD test cases and the closest training data descriptors provides improved correlation to prediction error of $R^2$ = 0.3 and $R^2$ = 0.2, respectively (Supporting Information Figure S30). Large errors (i.e., >15 kcal/mol) are only observed at a Euclidean norm difference exceeding 1.0 (half of the CSD data), providing an indication of lack of reliability in ANN prediction. This high distance to training data does not guarantee inaccurate prediction, e.g., CSD test case 8, a Fe(II) tetrapyridine complex, is predicted with fortuitously good ~2 kcal/mol error but has a Euclidean norm difference > 1.4. We have implemented the Euclidean norm metric alongside the ANN in our automated screening code[60] to detect complexes that are poorly represented in training data and advise retraining or direct calculation.

ANN-predicted equilibrium metal-ligand bond lengths for both HS and LS CSD geometries produced RMSEs of 0.10 and 0.07 Å, respectively (Supporting Information Tables S32-33). Trends in bond length prediction error differ from those obtained for spin-state splitting. For instance, bond length errors are average in the cyclams even though spin-state splitting predictions were poor. The large Euclidean distance to training data heuristic (> 1.0) is observed for five of the seven large (i.e., > 0.1 Å) HS bond distance errors (see Supporting Information Texts S4-5 and Figures S31-32). The highest HS prediction errors (>0.2 Å) occur for tests 8 and 35, underestimating the Fe-N bond length by 0.2 Å (2.1 Å ANN vs. 2.3 Å calculated) in the former case. Despite poor geometric predictions, the ANN predicts test 8 $\Delta E_{H-L}$ to within 3 kcal/mol, and this differing performance is due to the fact that predictions of these two outputs are independent. Interligand effects that are ignored by our descriptor set can restrict bond length extension, e.g. in test 16, where an O-H···O⁻ interligand hydrogen-bond produces an unusually short 1.9 Å high-spin Fe-N bond distance (vs. ANN prediction of 2.1 Å). Future work will focus on incorporating extended metrics of rigidity to account for these effects.

We investigated the relationship between the experimental CSD bond distances and the ANN-predicted bond distances. If the experimentally measured bond distance lies close to one spin state's predicted bond length, then the complex may be expected to be in that spin state, assuming i) the ANN provides a good prediction of the spin-state specific bond lengths and ii) that the gas-phase optimized DFT and CSD bond distances are comparable. The majority of experimental bond lengths are near the extrema of the ANN predictions (subset where ANN predicts LS-HS bond distance of at least 0.05 Å shown in Fig. 13). Nine of the twelve (9 of 9 in Fig. 13) experimental bond lengths that are on or above the predicted HS bond distance boundary have an HS ground state, eleven of the fifteen (6 of 6 in Fig. 13) experimental bond lengths that are on or below the predicted LS bond distance have an LS ground state, and remaining structures (3 in Fig. 13) reside at intermediate distances. Some discrepancies are due to differences between the gas phase geometries and those in the crystal environment (e.g., test 27 in Figure 13 and see Supporting Information Tables S31-33). This bond-length-based spin-assignment thus provides a strategy for corroboration of direct spin-state prediction.

**3.6 Extrapolating Pure Exchange-Correlation Functionals to Hybrids with an ANN**

Linear spin-state HF exchange sensitivity may be exploited to predict properties at one $a_{HF}$ value from computed properties obtained at another, e.g., to translate literature values or to accelerate periodic, plane-wave calculations where incorporation of HF exchange increases computational cost. We carry out comparison of the utility of this Δ-ML-inspired[45] strategy on the 35 CSD test set to identify if prediction errors are improved, especially for molecules poorly-represented in the training set.

On the CSD molecules, extrapolating $a_{HF}$=0.00 spin-state ordering to $a_{HF}$=0.20 with the exchange-sensitivity ANN reduces the maximum error to 23 kcal/mol and decrease the mean unsigned error and RMSE to 5 kcal/mol and 7 kcal/mol (the right pane of Figure 11 and Supporting Information Table S34). For the GGA + slope ANN approach, excluding the nine cyclams does not change the RMSE/MUE values, confirming good ANN exchange-sensitivity prediction even when spin-state splitting prediction is poor.

These reduced average errors are quite close to the uncertainty introduced by the slope prediction performance at around 4 kcal/mol over a 20% exchange interval. Although this approach does eliminate the largest outliers and improve prediction across the CSD test set, it necessitates semi-local DFT geometry optimizations or a judicious bond length choice for vertically-approximated spin-state ordering. This approach also has limited benefit for cases well-represented in the training data set due to the sparser data set in the exchange sensitivity ANN. Indeed, over the original test set molecules, extrapolated ANN exchange sensitivities on top of calculated $a_{HF} = 0.00$ splittings produce an RMSE of around 4 kcal/mol comparable to or slightly worse than direct prediction (Supporting Information Figure S33).

## 4. Conclusions

We have presented a series of ANN models trained using 2,690 DFT geometry optimizations of octahedral transition metal complexes generated from a set of 16 candidate axial and equatorial ligands and transition metals (Cr-Ni) at varying fractions of HF exchange. From the unseen test cases of a 60-40% train-test partition, we demonstrated good accuracy on spin-state splitting predictions of around 3 kcal/mol and metal-ligand bond distances around 0.02-0.03 Å. Our simple descriptor set, including: i) the ligand connection atom, ii) electronegativity and bonding of the coordinating ligand atom environment, iii) ligand formal charge, iv) ligand denticity, and v) metal identity and oxidation state ensures transferability of the ANN. Importantly, the employed connectivity models are not 3D-structure-based, instead relying on a truncated graph-theoretic representation of the ligand, making the approach suitable for screening large numbers of complexes without precise structural information. Although we have trained ANNs to predict bond lengths and spin-state splitting, the data set and descriptors could be used to predict other quantities such as ionization potential, redox potential, or molecular orbital energies. Such efforts are currently underway in our lab.

A test of our ANN on diverse molecules obtained from an experimental database indicated good performance, with MUEs of 5 kcal/mol for spin states for compounds within our proposed Euclidean distance reliability criteria and 10 kcal/mol for the full set. In both diverse and representative cases, the ANN outperforms other machine learning models. Our ANN predictions of HF exchange sensitivity provide a tool for interpolating between exchange-correlation functionals or extrapolating from semi-local GGAs to a hybrid result, which we demonstrated on CSD cases, improving MUE to 5 kcal/mol across the full 35 molecule set.

Natural extensions to this work include the development of the current ANN for extrapolation of GGA to hybrid functional properties in condensed matter systems and generalizing the coordination definition to enable prediction of properties of unsaturated metals in catalytic cycles. Overall, we have demonstrated a relatively sparse feature space to be capable of predicting electronic structure properties of transition metal complexes, and we anticipate that this strategy may be used for both high-throughput screening with knowledge of functional choice sensitivity and in guiding assessment of sources of errors in approximate DFT.

ASSOCIATED CONTENT

**Supporting Information**. Raw descriptor values and DFT splitting energies for test and training compounds; test and combined ANN prediction errors for splitting energy; raw descriptor values and DFT HFX slopes for test and training compounds; test and combined ANN prediction errors for HFX slopes; raw descriptor values and DFT low spin (LS) and high spin (HS) bond lengths for test and training compounds; test and combined ANN prediction errors for LS and HS bond lengths; ANN weight matrices (1,2,3), biases (1,2,3), and h2o configuration files for splitting, HFX slope and HS/LS bond length prediction; DFT optimized geometries for test and training data; DFT optimized geometries for CSD structures; excluded cases and reduced tolerances based geometry and spin contamination; ligand properties; LASSO variable selection coefficients and error analysis; descriptor scaling factors; hyperparameter estimation;

heteroleptic classification tree; spin splitting prediction errors for various metals and overall comparison of uncertainty and average error; HFX sensitivity by metal; HFX slope prediction errors for various metals and overall comparison of uncertainty and error; LS and HS bond length prediction errors for various metals and overall comparisons of uncertainty and error; ANN-assisted molSimplify performance; CSD structure details; CSD splitting energy predictions and comparison with similarity metrics; CSD bond length predictions and comparison with similarity metrics; comparison of interpolations by GGA and slope vs. direct prediction; ANN-predicted and calculated splitting energies, HFX slopes, and bond lengths.

## AUTHOR INFORMATION

**Corresponding Author**

*email: hjkulik@mit.edu phone: 617-253-4584

**Notes**

The authors declare no competing financial interest.

## ACKNOWLEDGMENT

The authors acknowledge partial support by the National Science Foundation under grant number ECCS-1449291. H.J.K. holds a Career Award at the Scientific Interface from the Burroughs Wellcome Fund. This work was carried out in part using computational resources from the Extreme Science and Engineering Discovery Environment (XSEDE), which is supported by National Science Foundation grant number ACI-1053575. The authors thank Adam H. Steeves for providing a critical reading of the manuscript and Prof. Youssef Marzouk for helpful conversations.

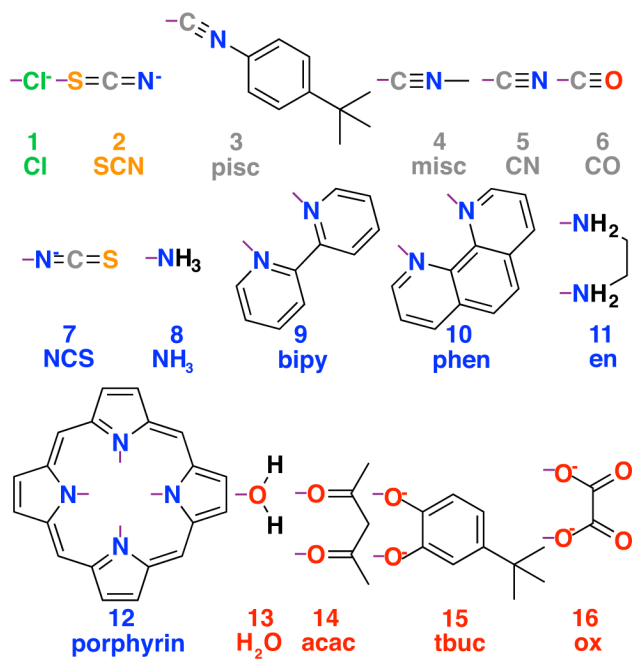

**Figure 1.** Set of ligands used to generate the transition metal complex data set. Ligands are numbered **1-16** and colored according to the atom type that coordinates with the metal, with chlorine in green, carbon in gray, sulfur in orange, nitrogen in blue, and oxygen in red. Purple lines indicate the bonds formed to metal-coordinating atoms in the ligand complexes. Abbreviations for each ligand used in the text are also shown. Full chemical names are provided in Supporting Information Table S3.

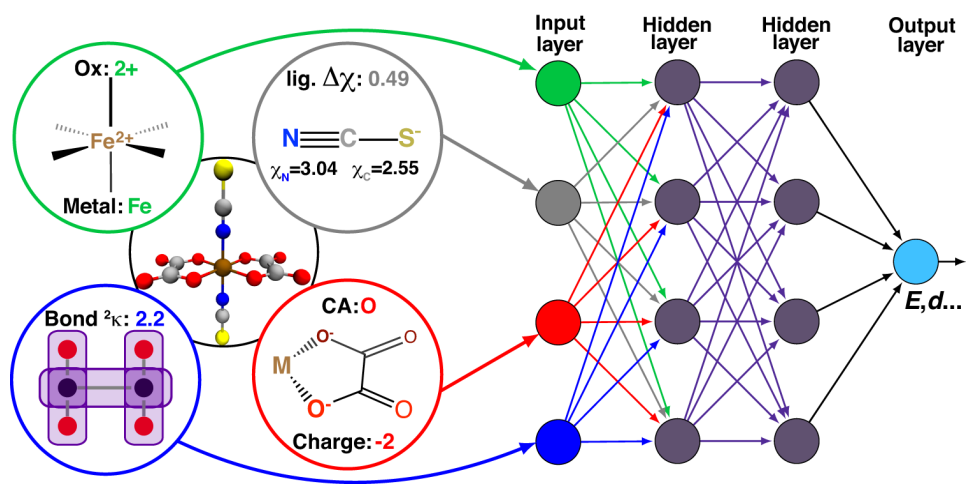

**Figure 2.** Schematic diagram of descriptors (left) as inputs to the ANN (right), along with hidden layers, and output (e.g., spin-state splittings) layers with additive bias term in each node omitted.

| Property | Variable set | | | | | | |
|---|---|---|---|---|---|---|---|
| | a | b | c | d | e | f | g |
| **Complex-based** | | | | | | | |
| Metal identity | ■ | ■ | ■ | ■ | ■ | ■ | ■ |
| Oxidation state | ■ | ■ | ■ | ■ | ■ | ■ | ■ |
| $a_{HF}$ | ■ | ■ | ■ | ■ | ■ | ■ | ■ |
| sum($\Delta\chi$) | | | ■ | | | | |
| min($\Delta\chi$) | | | ■ | | | | |
| max($\Delta\chi$) | | | ■ | ■ | ■ | ■ | ■ |
| **Ligand-based** | | | | | | | |
| Identity | ■ | | | | | | |
| Connection atom | | ■ | ■ | ■ | ■ | ■ | ■ |
| Charge | | ■ | ■ | ■ | ■ | ■ | ■ |
| Denticity | | ■ | ■ | ■ | ■ | ■ | ■ |
| Number of atoms | | ■ | ■ | ■ | ■ | | |
| Bond order | | | | | ■ | | ■ |
| Truncated Kier index | | | | | | ■ | ■ |

**Figure 3.** Summary of variables chosen for each set **a** through **g**. Employed variables are indicated in shaded gray and grouped by whether they are assessed on the whole complex (complex-based) or on each individual axial or equatorial ligand (ligand-based). $\Delta\chi$ is the difference in Pauling electronegativity between the ligand connecting atom and all atoms bonded to it, and the sum, maximum or minimum values are obtained over all ligands.

**Table 1.** Comparison of variable sets by root-mean-squared errors (RMSE) after regularization in $\Delta E_{H\text{-}L}$ and $\frac{\partial \Delta E_{H\text{-}L}}{\partial a_{HF}}$ prediction along with number of discrete variables (with all binary levels of the discrete variables counted in parentheses) and the number of continuous variables.

| set | RMSE($\Delta E_{H\text{-}L}$) (kcal/mol) | RMSE($\frac{\partial \Delta E_{H\text{-}L}}{\partial a_{HF}}$) (kcal/mol·HFX) | Discrete variables | Continuous variables |
|---|---|---|---|---|
| a | 14.6 | 20.6 | 3 (37) | 6 |
| b | 15.1 | 21.7 | 3 (15) | 8 |
| c | 15.2 | 21.2 | 3 (15) | 11 |
| d | 15.1 | 21.3 | 3 (15) | 10 |
| e | 14.9 | 21.1 | 3 (15) | 12 |
| f | 15.1 | 23.5 | 3 (15) | 10 |
| g | 14.9 | 21.3 | 3 (15) | 12 |

**Table 2.** Optimal (set **g**) input space descriptors and their range in the training set. $\Delta\chi$ is the difference in Pauling electronegativity between the ligand connecting atom and all atoms bonded to it. Here, a continuous descriptor corresponds to a single input node, whereas discrete descriptors correspond to one node per level.

| Symbol | Type | Descriptor | Values or Range |
|---|---|---|---|
| | | whole-complex descriptors | |
| M | Discrete | metal identity | Cr, Mn, Fe, Co, Ni |
| O | Continuous | oxidation state | 2 to 3 |
| me | Continuous | max. $\Delta\chi$ over all ligands | -0.89 to 1.20 |
| se | Continuous | sum of $\Delta\chi$ over all ligands | -5.30 to 7.20 |
| $a_{HF}$ | Continuous | HF exchange fraction | 0.00 to 0.30 |
| | | ligand-specific descriptors | |
| L | Discrete | ligand connection atom | Cl, S, C, N, or O |
| C | Continuous | ligand charge | 0 to -2 |
| k | Continuous | truncated Kier index | 0.00 to 6.95 |
| b | Continuous | ligand bond order | 0 to 3 |
| D | Continuous | ligand denticity | 1 to 4 |

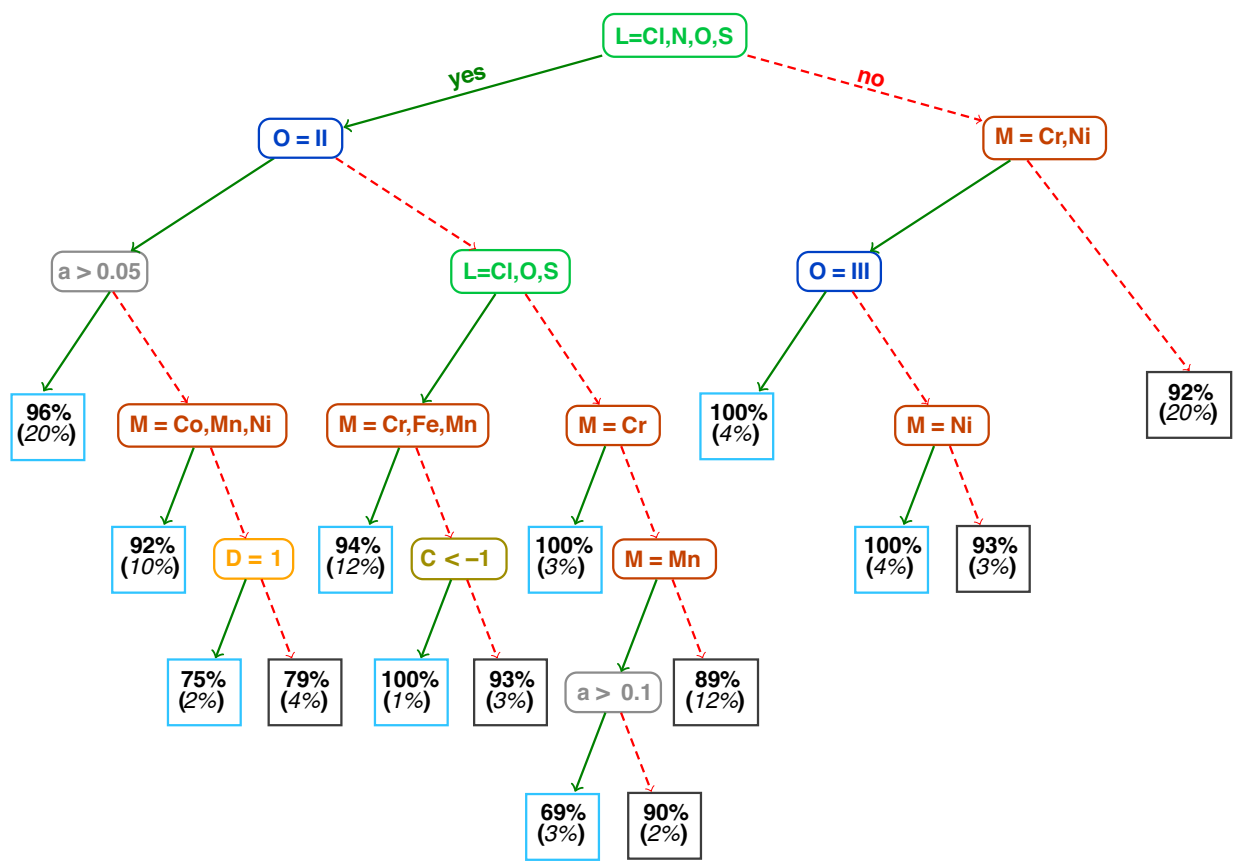

**Figure 4.** Binary ground state classification tree for homoleptic compounds. M indicates metal identity, L ligand connection atom, O oxidation state, a the fraction of HF exchange, C the charge, and D the ligand denticity. Each leaf node indicates the percent of elements in that leaf (light blue boxes for high-spin and dark gray boxes for low-spin) in bold font and percentage of total homoleptic population in the node (italic font, in parentheses).

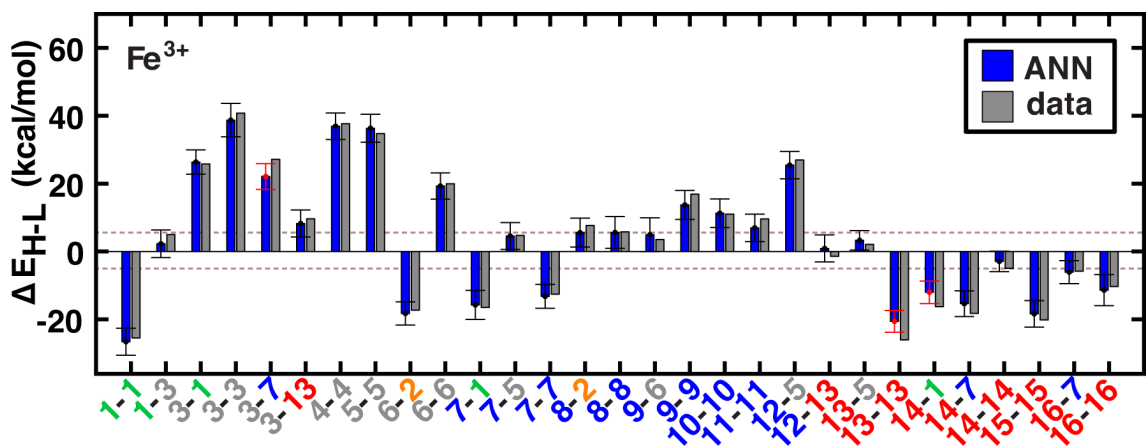

**Figure 5.** ANN model predictions (ANN, blue bars) and computed (data, gray bars) spin-state splittings, $\Delta E_{H-L}$, for the B3LYP functional ($a_{HF}$=0.20) in kcal/mol. Complexes are labeled by equatorial and then axial ligands according to the numbering indicated in Figure 1 and color-coded by direct ligand atom (green for chlorine, gray for carbon, blue for nitrogen, red for oxygen, and orange for sulfur). The error bars represent an estimated ±1 standard deviation credible interval from the mean prediction, and error bars that do not encompass the computed value are highlighted in red. Brown dashed lines correspond to a ±5 kcal/mol range around zero $\Delta E_{H-L}$, corresponding to near-degenerate spin states.

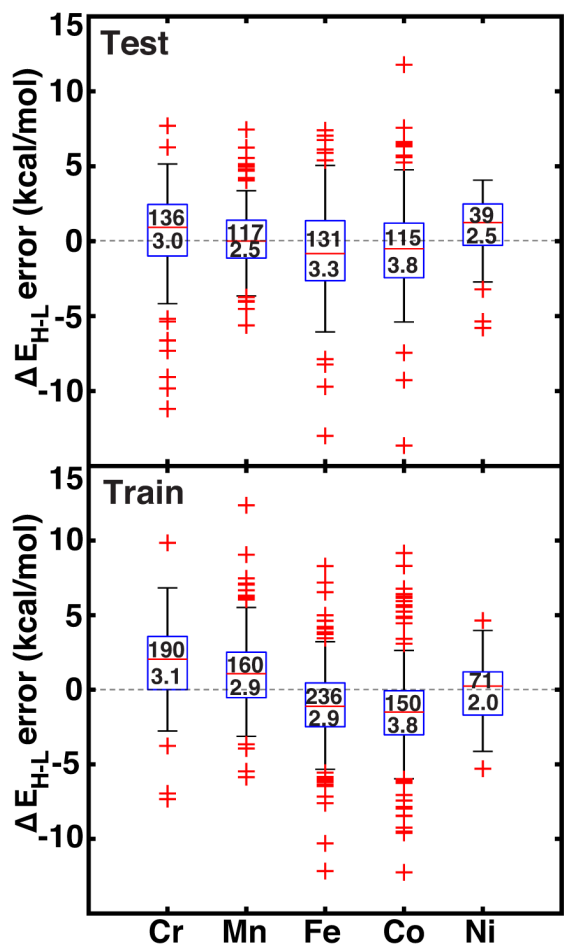

**Figure 6.** Error boxplots for $\Delta E_{H-L}$ in kcal/mol using the ANN for test (top) and training (bottom) data partitioned by metal identity. The top number inside the box indicates the number of cases in each set, and the bottom number indicates the RMSE in kcal/mol. The range for both graphs is from 15 kcal/mol to -15 kcal/mol.

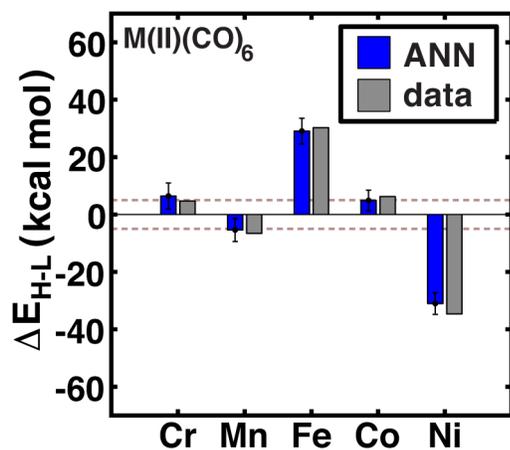

**Figure 7.** ANN model predictions (ANN, blue bars) and computed (data, gray bars) spin-state splittings, $\Delta E_{H-L}$, with the B3LYP functional ($a_{HF}$=0.20) in kcal/mol on $M(II)(CO)_6$ complexes, where M = Cr, Mn, Fe, Co, or Ni. The error bars represent an estimated ±1 standard deviation credible interval from the mean prediction, and brown dashed lines correspond to a ±5 kcal/mol range around zero $\Delta E_{H-L}$, corresponding to near-degenerate spin states.

**Table 3.** Train/test data and CSD test set RMSEs and max UEs in kcal/mol·HFX$^{-1}$ for different machine learning methods and descriptor sets compared: KRR, kernel ridge regression, using square-exponential kernel for descriptor set g and the L1 matrix distance[45] for the sorted Coulomb matrix descriptor; SVR, support vector regression using square-exponential kernel; ANN, artificial neural network. Results are also given for the KRR/Coulomb case, restricted to B3LYP only since the Coulomb matrix does not naturally account for varying HF exchange.

| Model | Descriptor | Training | | Test | | CSD | |
|---|---|---|---|---|---|---|---|
| | | RMSE | max UE | RMSE | max UE | RMSE | max UE |
| LASSO | set g | 16.1 | 89.7 | 15.7 | 93.5 | 19.2 | 72.5 |
| KRR | set g | 1.6 | 8.5 | 3.9 | 17.0 | 38.3 | 88.4 |
| SVR | set g | 2.1 | 20.9 | 3.6 | 20.4 | 20.3 | 64.8 |
| ANN | set g | 3.0 | 12.3 | 3.1 | 15.6 | 13.1 | 30.4 |
| KRR | sorted Coulomb | 4.3 | 41.5 | 30.8 | 103.7 | 54.5 | 123.9 |
| KRR, B3LYP only | sorted Coulomb | 17.2 | 58.0 | 28.1 | 69.5 | 46.7 | 118.7 |

**Table 4.** Test set RMSEs in kcal/mol·HFX$^{-1}$ separated by metal and oxidation state along with minimum and maximum unsigned test errors (UE). The number of test cases is indicated in parentheses.

| Species | RMSE | min. UE | max. UE |
|---|---|---|---|
| Cr(II) | 21 (*14*) | 4 | 45 |
| Cr(III) | 17 (*8*) | 2 | 37 |
| Mn(II) | 24 (*6*) | 3 | 40 |
| Mn(III) | 38 (*8*) | 4 | 92 |
| Fe(II) | 18 (*9*) | 2 | 41 |
| Fe(III) | 15 (*12*) | <1 | 32 |
| Co(II) | 17 (*8*) | <1 | 26 |
| Co(III) | 20 (*8*) | <1 | 46 |
| Ni(II) | 9 (*4*) | 1 | 15 |

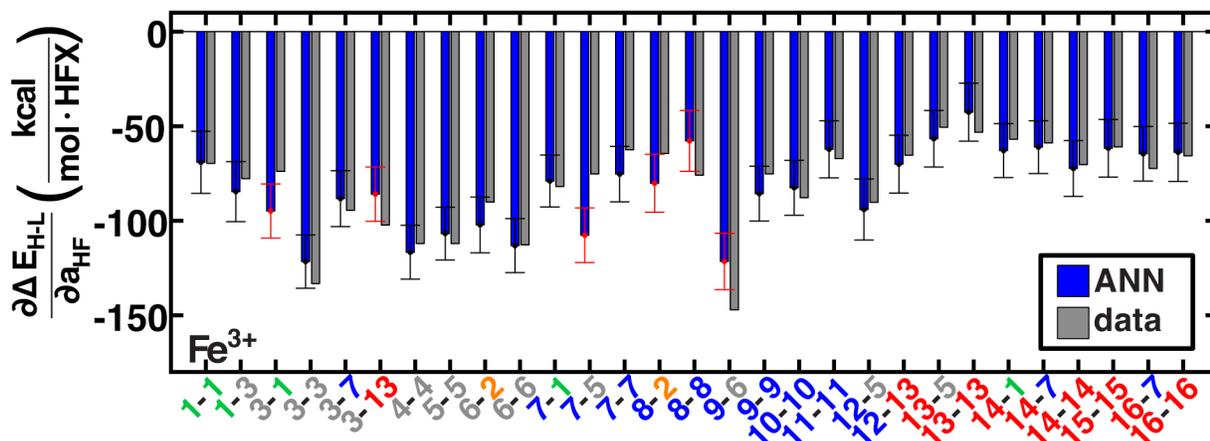

**Figure 8.** ANN model predictions (ANN, blue bars) and computed (data, gray bars) spin-state splitting sensitivities to HF exchange, $\frac{\partial \Delta E_{H-L}}{\partial a_{HF}}$, in kcal/mol·HFX$^{-1}$, for $Fe^{3+}$ complexes. Complexes are labeled as equatorial and then axial ligands according to the numbering indicated in Figure 1 and color-coded by direct ligand atom (green for chlorine, gray for carbon, blue for nitrogen, red for oxygen, and orange for sulfur). The error bars represent an estimated ±1 standard deviation credible interval from the mean prediction, and error bars that do not encompass the computed value are highlighted in red.

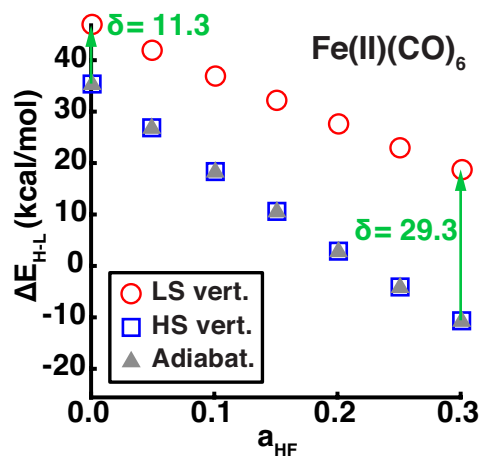

**Figure 9.** The vertical or adiabatic spin-state splittings, $\Delta E_{H-L}$, in kcal/mol as a function of HF exchange, $a_{HF}$, for $Fe(II)(CO)_6$. Spin-state splittings evaluated at the HS or LS geometries are indicated by open blue squares and open red circles, respectively. The adiabatic spin-state splitting is shown as filled gray triangles. The HS vertical and adiabatic splittings overlap, whereas the LS vertical splitting overestimates $\Delta E_{H-L}$, as indicated by the green arrow and annotated δ in kcal/mol for $a_{HF}$=0.00 and $a_{HF}$=0.30.

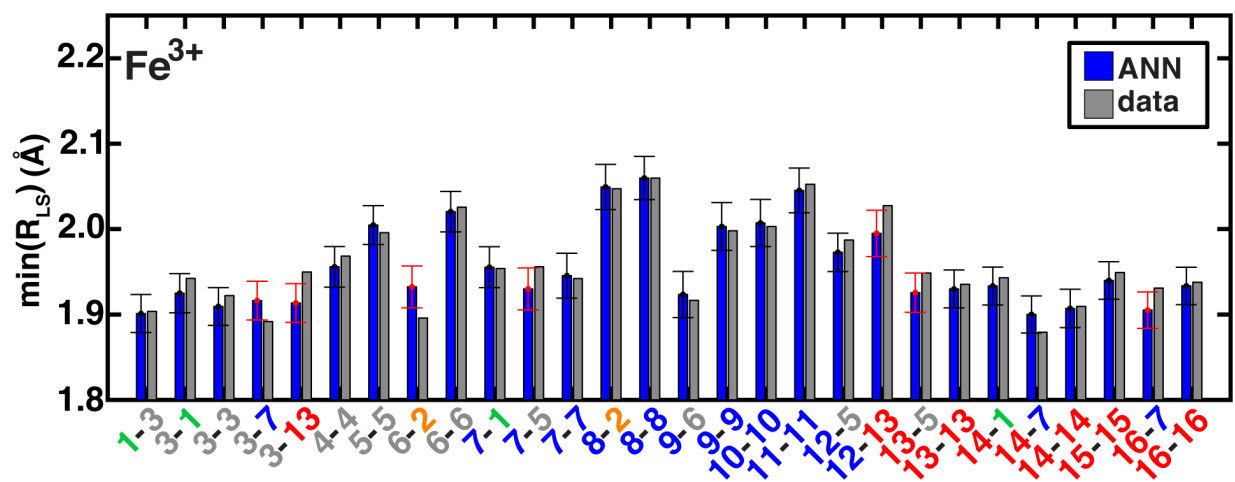

**Figure 10.** ANN model predictions (ANN, blue bars) and computed (data, gray bars) minimum LS $Fe^{3+}$ bond lengths, min($R_{LS}$), in Å. Complexes are labeled as equatorial and then axial ligands according to the numbering indicated in Figure 1 and color-coded by direct ligand atom (green for chlorine, gray for carbon, blue for nitrogen, and red for oxygen). The error bars represent an estimated ±1 standard deviation credible interval around the mean prediction, and error bars that do not encompass the computed value are highlighted in red. Fe(III)(Cl)$_6$ (**1-1**) is excluded due to being off scale: it has a predicted/calculated bond length of 2.44/2.45 Å, and an error standard deviation of ±0.02.

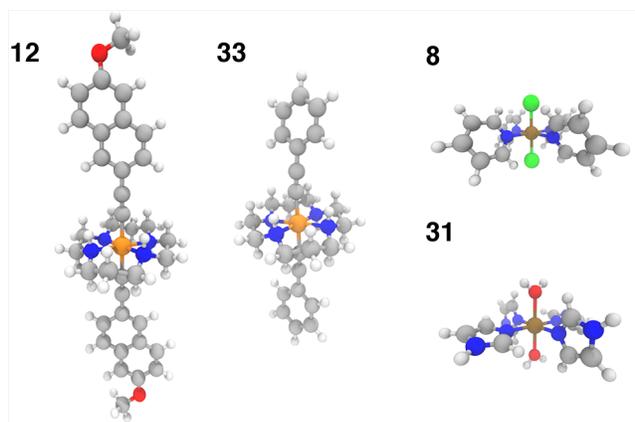

**Figure 11.** Representative CSD test set molecules shown in ball and stick representation with carbon atoms in gray, nitrogen atoms in blue, oxygen in red, hydrogen in white, chlorine in green, chromium in orange, and iron in brown. Test molecules 12 (CSD ID: SUMLET) and 33 (CSD ID: YUJCIQ) are Cr(III) cyclams for which the ANN performs least well, and test molecules 8 (CSD ID: TPYFEC04) and 31 (CSD ID: BIPGEN) are cases for which the ANN predicts $\Delta E_{H\text{-}L}$ within 3 kcal/mol.

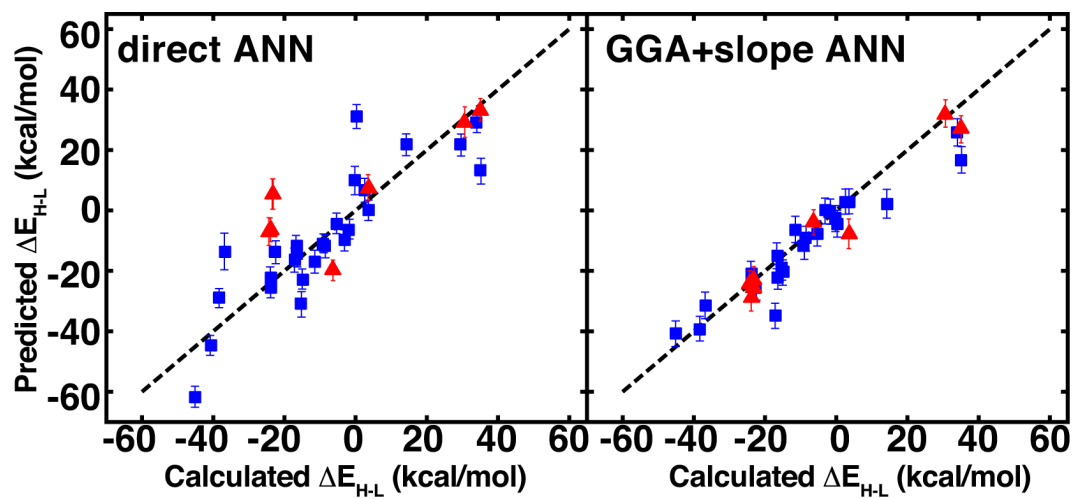

**Figure 12**. ANN spin-state splitting energy, $\Delta E_{H-L}$, prediction on CSD test structures vs. DFT-calculated values, both at $a_{HF}$ = 0.20 and in kcal/mol. Direct prediction (left) is compared to GGA calculations and extrapolation using the predicted slope from the ANN (right). Error bars represent a credible interval of one standard deviation from the model uncertainty analysis (either in direct ANN at left or slope ANN at right), and a parity line (black, dashed) is indicated. Cyclams are indicated in red triangles, as described in main text, and the remaining test cases are indicated by blue squares.

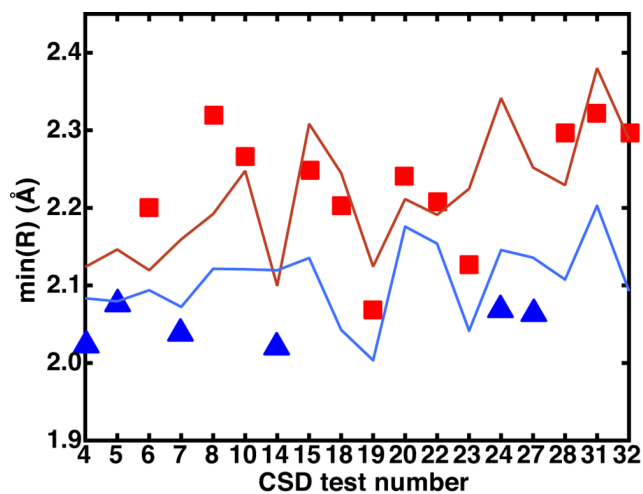

**Figure 13.** Comparison of measured CSD bond distances in the crystal phase, represented by symbols (red squares for high-spin or blue triangles for low-spin based on DFT assignment at $a_{HF}$=0.20) with the ANN predicted HS (red line) and LS (blue line) bond distances. Only the CSD test cases where the difference between ANN-predicted LS and HS bond distances is ≥ 0.05 Å are shown for clarity. For all of these cases, the ANN correctly predicts the DFT spin state.

**REFERENCES**

1. Gomez-Bombarelli, R.; Aguilera-Iparraguirre, J.; Hirzel, T. D.; Duvenaud, D.; Maclaurin, D.; Blood-Forsythe, M. A.; Chae, H. S.; Einzinger, M.; Ha, D. G.; Wu, T.; Markopoulos, G.; Jeon, S.; Kang, H.; Miyazaki, H.; Numata, M.; Kim, S.; Huang, W.; Hong, S. I.; Baldo, M.; Adams, R. P.; Aspuru-Guzik, A., Design of Efficient Molecular Organic Light-Emitting Diodes by a High-Throughput Virtual Screening and Experimental Approach. *Nat. Mater.* **2016,** *15*, 1120-1127.
2. Pyzer-Knapp, E. O.; Li, K.; Aspuru-Guzik, A., Learning from the Harvard Clean Energy Project: The Use of Neural Networks to Accelerate Materials Discovery. *Adv. Funct. Mater.* **2015,** *25*, 6495-6502.
3. Norskov, J. K.; Bligaard, T., The Catalyst Genome. *Angew. Chem., Int. Ed. Engl.* **2013,** *52*, 776-7.
4. Jain, A.; Ong, S. P.; Hautier, G.; Chen, W.; Richards, W. D.; Dacek, S.; Cholia, S.; Gunter, D.; Skinner, D.; Ceder, G., Commentary: The Materials Project: A Materials Genome Approach to Accelerating Materials Innovation. *APL Mater.* **2013,** *1*, 011002.
5. Virshup, A. M.; Contreras-García, J.; Wipf, P.; Yang, W.; Beratan, D. N., Stochastic Voyages into Uncharted Chemical Space Produce a Representative Library of All Possible Drug-Like Compounds. *J. Am. Chem. Soc.* **2013,** *135*, 7296-7303.
6. Kirkpatrick, P.; Ellis, C., Chemical Space. *Nature* **2004,** *432*, 823-823.
7. Meredig, B.; Agrawal, A.; Kirklin, S.; Saal, J. E.; Doak, J. W.; Thompson, A.; Zhang, K.; Choudhary, A.; Wolverton, C., Combinatorial Screening for New Materials in Unconstrained Composition Space with Machine Learning. *Phys. Rev. B* **2014,** *89*, 094104.
8. Li, L.; Snyder, J. C.; Pelaschier, I. M.; Huang, J.; Niranjan, U.-N.; Duncan, P.; Rupp, M.; Müller, K.-R.; Burke, K., Understanding Machine-Learned Density Functionals. *Int. J. Quantum Chem.* **2016,** *116*, 819-833.
9. Rupp, M., Machine Learning for Quantum Mechanics in a Nutshell. *Int. J. Quantum Chem.* **2015,** *115*, 1058-1073.
10. Behler, J., Perspective: Machine Learning Potentials for Atomistic Simulations. *J. Chem. Phys.* **2016,** *145*, 170901.
11. Behler, J., Representing Potential Energy Surfaces by High-Dimensional Neural Network Potentials. *J. Phys.: Condens. Matter* **2014,** *26*, 183001.
12. Lorenz, S.; Groß, A.; Scheffler, M., Representing High-Dimensional Potential-Energy Surfaces for Reactions at Surfaces by Neural Networks. *Chem. Phys. Lett.* **2004,** *395*, 210-215.
13. Artrith, N.; Morawietz, T.; Behler, J., High-Dimensional Neural-Network Potentials for Multicomponent Systems: Applications to Zinc Oxide. *Phys. Rev. B* **2011,** *83*, 153101.
14. Behler, J.; Parrinello, M., Generalized Neural-Network Representation of High-Dimensional Potential-Energy Surfaces. *Phys. Rev. Lett.* **2007,** *98*, 146401.
15. Prudente, F. V.; Neto, J. J. S., The Fitting of Potential Energy Surfaces Using Neural Networks. Application to the Study of the Photodissociation Processes. *Chem. Phys. Lett.* **1998,** *287*, 585-589.
16. Mones, L.; Bernstein, N.; Csanyi, G., Exploration, Sampling, and Reconstruction of Free Energy Surfaces with Gaussian Process Regression. *J. Chem. Theory Comput.* **2016,** *12*, 5100-5110.
17. Snyder, J. C.; Rupp, M.; Hansen, K.; Muller, K.-R.; Burke, K., Finding Density Functionals with Machine Learning. *Phys. Rev. Lett.* **2012,** *108*, 253002.

18. Yao, K.; Parkhill, J., Kinetic Energy of Hydrocarbons as a Function of Electron Density and Convolutional Neural Networks. *J. Chem. Theory Comput.* **2016,** *12*, 1139-1147.
19. Snyder, J. C.; Rupp, M.; Hansen, K.; Blooston, L.; Müller, K.-R.; Burke, K., Orbital-Free Bond Breaking via Machine Learning. *J. Chem. Phys.* **2013,** *139*, 224104.
20. Hase, F.; Valleau, S.; Pyzer-Knapp, E.; Aspuru-Guzik, A., Machine Learning Exciton Dynamics. *Chem. Sci.* **2016,** *7*, 5139-5147.
21. Li, Z.; Kermode, J. R.; De Vita, A., Molecular Dynamics with on-the-Fly Machine Learning of Quantum-Mechanical Forces. *Phys. Rev. Lett.* **2015,** *114*, 096405.
22. Botu, V.; Ramprasad, R., Adaptive Machine Learning Framework to Accelerate Ab Initio Molecular Dynamics. *Int. J. Quantum Chem.* **2015,** *115*, 1074-1083.
23. Ma, X.; Li, Z.; Achenie, L. E. K.; Xin, H., Machine-Learning-Augmented Chemisorption Model for CO2 Electroreduction Catalyst Screening. *J. Phys. Chem. Lett.* **2015,** *6*, 3528-3533.
24. Mannodi-Kanakkithodi, A.; Pilania, G.; Huan, T. D.; Lookman, T.; Ramprasad, R., Machine Learning Strategy for Accelerated Design of Polymer Dielectrics. *Sci. Rep.* **2016,** *6*, 20952.
25. Huan, T. D.; Mannodi-Kanakkithodi, A.; Ramprasad, R., Accelerated Materials Property Predictions and Design Using Motif-Based Fingerprints. *Phys. Rev. B* **2015,** *92*, 014106.
26. Pilania, G.; Wang, C.; Jiang, X.; Rajasekaran, S.; Ramprasad, R., Accelerating Materials Property Predictions Using Machine Learning. *Sci. Rep.* **2013,** *3*, 2810.
27. Lee, J.; Seko, A.; Shitara, K.; Tanaka, I., Prediction Model of Band-Gap for AX Binary Compounds by Combination of Density Functional Theory Calculations and Machine Learning Techniques. *Phys. Rev. B* **2016,** *93*, 115104.
28. Morawietz, T.; Behler, J., A Density-Functional Theory-Based Neural Network Potential for Water Clusters Including Van Der Waals Corrections. *J. Phys. Chem. A* **2013,** *117*, 7356-7366.
29. Morawietz, T.; Singraber, A.; Dellago, C.; Behler, J., How Van Der Waals Interactions Determine the Unique Properties of Water. *Proc. Natl. Acad. Sci. U. S. A.* **2016**, 201602375.
30. Rupp, M.; Tkatchenko, A.; Müller, K.-R.; von Lilienfeld, O. A., Fast and Accurate Modeling of Molecular Atomization Energies with Machine Learning. *Phys. Rev. Lett.* **2012,** *108*, 058301.
31. Huang, B.; von Lilienfeld, O. A., Communication: Understanding Molecular Representations in Machine Learning: The Role of Uniqueness and Target Similarity. *J. Chem. Phys.* **2016,** *145*, 161102.
32. De, S.; Bartk, A. P.; Csányi, G.; Ceriotti, M., Comparing Molecules and Solids across Structural and Alchemical Space. *Phys. Chem. Chem. Phys.* **2016,** *18*, 1-18.
33. Maggiora, G.; Vogt, M.; Stumpfe, D.; Bajorath, J., Molecular Similarity in Medicinal Chemistry: Miniperspective. *J. Med. Chem.* **2013,** *57*, 3186-3204.
34. Wang, J.; Wolf, R. M.; Caldwell, J. W.; Kollman, P. A.; Case, D. A., Development and Testing of a General Amber Force Field. *J. Comp. Chem.* **2004,** *25*, 1157-1174.
35. Kubinyi, H., QSAR and 3D QSAR in Drug Design. Part 1: Methodology. *Drug Discovery Today* **1997,** *2*, 457-467.
36. Benson, S. W.; Cruickshank, F.; Golden, D.; Haugen, G. R.; O'neal, H.; Rodgers, A.; Shaw, R.; Walsh, R., Additivity Rules for the Estimation of Thermochemical Properties. *Chem. Rev.* **1969,** *69*, 279-324.
37. Deeth, R. J., The Ligand Field Molecular Mechanics Model and the Stereoelectronic Effects of d and S Electrons. *Coord. Chem. Rev.* **2001,** *212*, 11-34.

38. Shriver, D. F.; Atkins, P. W., *Inorganic Chemistry*. 3rd ed.; W.H. Freeman and Co.: 1999.
39. Schütt, K. T.; Glawe, H.; Brockherde, F.; Sanna, A.; Müller, K. R.; Gross, E. K. U., How to Represent Crystal Structures for Machine Learning: Towards Fast Prediction of Electronic Properties. *Phys. Rev. B* **2014,** *89*, 205118.
40. Ioannidis, E. I.; Kulik, H. J., Towards Quantifying the Role of Exact Exchange in Predictions of Transition Metal Complex Properties. *J. Chem. Phys.* **2015,** *143*, 034104.
41. Gani, T. Z. H.; Kulik, H. J., Where Does the Density Localize? Convergent Behavior for Global Hybrids, Range Separation, and DFT+U *arxiv preprint arXiv:1610.01222* **2016**.
42. Ioannidis, E. I.; Kulik, H. J., Ligand-Field-Dependent Behavior of Meta-GGA Exchange in Transition-Metal Complex Spin-State Ordering. *J. Phys. Chem. A* **2017,** *just accepted*.
43. Huang, W.; Xing, D.-H.; Lu, J.-B.; Long, B.; Schwarz, W. E.; Li, J., How Much Can Density Functional Approximations (DFA) Fail? The Extreme Case of the FeO4 Species. *J. Chem. Theory Comput.* **2016,** *12*, 1525-1533.
44. Stewart, J. J. P., Optimization of Parameters for Semiempirical Methods VI: More Modifications to the NDDO Approximations and Re-Optimization of Parameters. *J. Mol. Model.* **2013,** *19*, 1-32.
45. Ramakrishnan, R.; Dral, P. O.; Rupp, M.; von Lilienfeld, O. A., Big Data Meets Quantum Chemistry Approximations: The Delta-Machine Learning Approach. *J. Chem. Theory Comput.* **2015,** *11*, 2087-96.
46. Shen, L.; Wu, J.; Yang, W., Multiscale Quantum Mechanics/Molecular Mechanics Simulations with Neural Networks. *J. Chem. Theory Comput.* **2016,** *12*, 4934-4946.
47. Kulik, H. J., Perspective: Treating Electron over-Delocalization with the DFT+U Method. *The Journal of Chemical Physics* **2015,** *142*, 240901.
48. Cohen, A. J.; Mori-Sanchez, P.; Yang, W., Insights into Current Limitations of Density Functional Theory. *Science* **2008,** *321*, 792-794.
49. Salomon, O.; Reiher, M.; Hess, B. A., Assertion and Validation of the Performance of the B3LYP* Functional for the First Transition Metal Row and the G2 Test Set. *J. Chem. Phys.* **2002,** *117*, 4729-4737.
50. Reiher, M., Theoretical Study of the Fe(Phen)2(NCS)2 Spin-Crossover Complex with Reparametrized Density Functionals. *Inorg. Chem.* **2002,** *41*, 6928-6935.
51. Reiher, M.; Salomon, O.; Hess, B. A., Reparameterization of Hybrid Functionals Based on Energy Differences of States of Different Multiplicity. *Theor. Chem. Acc.* **2001,** *107*, 48-55.
52. Droghetti, A.; Alfè, D.; Sanvito, S., Assessment of Density Functional Theory for Iron(II) Molecules across the Spin-Crossover Transition. *J. Chem. Phys.* **2012,** *137*, 124303.
53. Sutton, J. E.; Guo, W.; Katsoulakis, M. A.; Vlachos, D. G., Effects of Correlated Parameters and Uncertainty in Electronic-Structure-Based Chemical Kinetic Modelling. *Nat. Chem.* **2016,** *8*, 331-337.
54. Simm, G. N.; Reiher, M., Systematic Error Estimation for Chemical Reaction Energies. *J. Chem. Theory Comput.* **2016**.
55. Walker, E.; Ammal, S. C.; Terejanu, G. A.; Heyden, A., Uncertainty Quantification Framework Applied to the Water–Gas Shift Reaction over Pt-Based Catalysts. *The Journal of Physical Chemistry C* **2016,** *120*, 10328-10339.
56. Halcrow, M. A., Structure: Function Relationships in Molecular Spin-Crossover Complexes. *Chem. Soc. Rev.* **2011,** *40*, 4119-4142.

57.     Létard, J.-F.; Guionneau, P.; Goux-Capes, L., Towards Spin Crossover Applications. In *Spin Crossover in Transition Metal Compounds III*, Springer: 2004; pp 221-249.
58.     Bignozzi, C. A.; Argazzi, R.; Boaretto, R.; Busatto, E.; Carli, S.; Ronconi, F.; Caramori, S., The Role of Transition Metal Complexes in Dye-Sensitized Solar Devices. *Coord. Chem. Rev.* **2013,** *257*, 1472-1492.
59.     Harvey, J. N.; Poli, R.; Smith, K. M., Understanding the Reactivity of Transition Metal Complexes Involving Multiple Spin States. *Coord. Chem. Rev.* **2003,** *238*, 347-361.
60.     Ioannidis, E. I.; Gani, T. Z. H.; Kulik, H. J., Molsimplify: A Toolkit for Automating Discovery in Inorganic Chemistry. *J. Comp. Chem.* **2016,** *37*, 2106-2117.
61.     Kramida, A., Ralchenko, Yu., Reader, J. and NIST ASD Team NIST Atomic Spectra Database (Version 5.3). http://physics.nist.gov/asd (accessed November 5, 2016).
62.     Groom, C. R.; Bruno, I. J.; Lightfoot, M. P.; Ward, S. C., The Cambridge Structural Database. *Acta Crystallogr., Sect. B: Struct. Sci., Cryst. Eng. Mater.* **2016,** *72*, 171-179.
63.     Ufimtsev, I. S.; Martinez, T. J., Quantum Chemistry on Graphical Processing Units. 3. Analytical Energy Gradients, Geometry Optimization, and First Principles Molecular Dynamics. *J. Chem. Theory Comput.* **2009,** *5*, 2619-2628.
64.     Petachem. http://www.petachem.com. (accessed November 5, 2016).
65.     Stephens, P. J.; Devlin, F. J.; Chabalowski, C. F.; Frisch, M. J., Ab Initio Calculation of Vibrational Absorption and Circular Dichroism Spectra Using Density Functional Force Fields. *J. Phys. Chem.* **1994,** *98*, 11623-11627.
66.     Becke, A. D., Density-Functional Thermochemistry. III. The Role of Exact Exchange. *J. Chem. Phys.* **1993,** *98*, 5648-5652.
67.     Lee, C.; Yang, W.; Parr, R. G., Development of the Colle-Salvetti Correlation-Energy Formula into a Functional of the Electron Density. *Phys. Rev. B* **1988,** *37*, 785-789.
68.     Hay, P. J.; Wadt, W. R., Ab Initio Effective Core Potentials for Molecular Calculations. Potentials for the Transition Metal Atoms Sc to Hg. *J. Chem. Phys.* **1985,** *82*, 270-283.
69.     Saunders, V. R.; Hillier, I. H., A "Level–Shifting" Method for Converging Closed Shell Hartree–Fock Wave Functions. *Int. J. Quantum Chem.* **1973,** *7*, 699-705.
70.     Kästner, J.; Carr, J. M.; Keal, T. W.; Thiel, W.; Wander, A.; Sherwood, P., DL-FIND: An Open-Source Geometry Optimizer for Atomistic Simulations. *J. Phys. Chem. A* **2009,** *113*, 11856-11865.
71.     Ganzenmuller, G.; Berkaine, N.; Fouqueau, A.; Casida, M. E.; Reiher, M., Comparison of Density Functionals for Differences between the High- (5t2g) and Low- (1a1g) Spin States of Iron(II) Compounds. IV. Results for the Ferrous Complexes [Fe(L)('NHS4')]. *J. Chem. Phys.* **2005,** *122*, 234321.
72.     Cereto-Massague, A.; Ojeda, M. J.; Valls, C.; Mulero, M.; Garcia-Vallve, S.; Pujadas, G., Molecular Fingerprint Similarity Search in Virtual Screening. *Methods* **2015,** *71*, 58-63.
73.     Sheridan, R. P.; Miller, M. D.; Underwood, D. J.; Kearsley, S. K., Chemical Similarity Using Geometric Atom Pair Descriptors. *J. Chem. Inf. Model.* **1996,** *36*, 128-136.
74.     Hansen, K.; Biegler, F.; Ramakrishnan, R.; Pronobis, W., Machine Learning Predictions of Molecular Properties: Accurate Many-Body Potentials and Nonlocality in Chemical Space. *J. Phys. Chem. Lett.* **2015,** *6*, 2326-2331.
75.     Gastegger, M.; Marquetand, P., High-Dimensional Neural Network Potentials for Organic Reactions and an Improved Training Algorithm. *J. Chem. Theory Comput.* **2015,** *11*, 2187-2198.

76. Hageman, J. A.; Westerhuis, J. A.; Frhauf, H. W.; Rothenberg, G., Design and Assembly of Virtual Homogeneous Catalyst Libraries - Towards in Silico Catalyst Optimisation. *Adv. Synth. Catal.* **2006,** *348*, 361-369.
77. Randic, M., On Molecular Branching. *J. Am. Chem. Soc.* **1975,** *97*, 57-77.
78. Wiener, H., Correlation of Heats of Isomerization, and Differences in Heats of Vaporization of Isomers, among the Paraffin Hydrocarbons. *J. Am. Chem. Soc.* **1947,** *69*, 2636-2638.
79. Kier, L. B., A Shape Index from Molecular Graphs. *Quant. Struct.-Act. Relat.* **1985,** *4*, 109-116.
80. Montavon, G.; Hansen, K.; Fazli, S.; Rupp, M. In *Learning Invariant Representations of Molecules for Atomization Energy Prediction*, Advances in Neural Information Processing Systems, Pereira, F.; Burges, C. J. C.; Bottou, L.; Weinberger, K. Q., Eds. Curran Associates, Inc.: 2012; pp 440-448.
81. Gastegger, M.; Kauffmann, C.; Behler, J.; Marquetand, P., Comparing the Accuracy of High-Dimensional Neural Network Potentials and the Systematic Molecular Fragmentation Method: A Benchmark Study for All-Trans Alkanes. *J. Chem. Phys.* **2016,** *144*, 194110.
82. Hastie, T.; Tibshirani, R.; Friedman, J., *The Elements of Statistical Learning*. Springer New York: 2009; Vol. 18, p 764.
83. Friedman, J.; Hastie, T.; Tibshirani, R., Regularization Paths for Generalized Linear Models via Coordinate Descent. *J. Stat. Softw.* **2010,** *33*, 1-22.
84. R Core Development Team, R: A Language and Environment for Statistical Computing. 2016.
85. Larochelle, H.; Bengio, Y.; Louradour, J.; Lamblin, P., Exploring Strategies for Training Deep Neural Networks *J. Mach. Learn. Res.* **2009,** *10*, 1-40.
86. Aiello, S.; Kraljevic, T.; Maj, P. *H2O: R Interface for H2O*.; 2015.
87. Gal, Y.; Ghahramani, Z., Dropout as a Bayesian Approximation: Representing Model Uncertainty in Deep Learning. *arXiv preprint arxiv:1506.02142* **2015**.
88. Srivastava, N.; Hinton, G. E.; Krizhevsky, A.; Sutskever, I.; Salakhutdinov, R., Dropout : A Simple Way to Prevent Neural Networks from Overfitting. *J. Mach. Learn. Res.* **2014,** *15*, 1929-1958.
89. Hinton, G. E.; Srivastava, N.; Krizhevsky, A.; Sutskever, I.; Salakhutdinov, R. R., Improving Neural Networks by Preventing Co-Adaptation of Feature Detectors. *arXiv preprint arXiv:1207.0580* **2012**, 1-18.
90. Bengio, Y., Practical Recommendations for Gradient-Based Training of Deep Architectures. In *Neural Networks: Tricks of the Trade*, Orr, G. B.; Muller, K.-R.; Gregoire, M., Eds. Springer: 2012; pp 437-478.
91. LeCun, Y.; Bottou, L.; Bengio, Y.; Haffner, P., Gradient-Based Learning Applied to Document Recognition. *Proc. IEEE* **1998,** *86*, 2278-2323.
92. Candel, A.; Parmar, V.; LeDell, E.; Arora, A., Deep Learning with H2O. H2O: 2015.
93. Niu, F.; Recht, B.; Re, C.; Wright, S. J., Hogwild!: A Lock-Free Approach to Parallelizing Stochastic Gradient Descent. *Advances in Neural Information Processing Systems* **2011**, 21.
94. Kingston, G. B.; Lambert, M. F.; Maier, H. R., Bayesian Training of Artificial Neural Networks Used for Water Resources Modeling. *Water Resour. Res.* **2005,** *41*, 1-11.

95. Secchi, P.; Zio, E. a., Quantifying Uncertainties in the Estimation of Safety Parameters by Using Bootstrapped Artificial Neural Networks. *Annals of Nuclear Energy* **2008,** *35*, 2338-2350.
96. Hansen, K.; Montavon, G.; Biegler, F.; Fazli, S.; Rupp, M.; Scheffler, M.; von Lilienfeld, O. A.; Tkatchenko, A.; Müller, K.-R., Assessment and Validation of Machine Learning Methods for Predicting Molecular Atomization Energies. *J. Chem. Theory Comput.* **2013,** *9*, 3404-3419.
97. Zeileis, A.; Hornik, K.; Smola, A.; Karatzoglou, A., Kernlab - an S4 Package for Kernel Methods in R. *J. Stat. Softw.* **2004,** *11*, 1-20.
98. Krueger, T.; Panknin, D.; Braun, M., Fast Cross-Validation via Sequential Testing. *J. Mach. Learn. Res.* **2015,** *16*, 1103-1155.
99. Breiman, L.; Friedman, J.; Olshen, R. A.; Stone, C., *Classification and Regression Trees*. Chapman and Hall/CRC: 1984; Vol. 5, p 95-96.
100. Therneau, T.; Atkinson, B.; Ripley, B. Rpart: Recursive Partitioning and Regression Trees. https://cran.r-project.org/package=rpart (accessed November 5, 2016).
101. Coskun, D.; Jerome, S. V.; Friesner, R. A., Evaluation of the Performance of the B3LYP, PBE0, and M06 DFT Functionals, and DBLOC-Corrected Versions, in the Calculation of Redox Potentials and Spin Splittings for Transition Metal Containing Systems. *J. Chem. Theory Comput.* **2016,** *12*, 1121-8.
102. Hughes, T. F.; Harvey, J. N.; Friesner, R. A., A B3LYP-DBLOC Empirical Correction Scheme for Ligand Removal Enthalpies of Transition Metal Complexes: Parameterization against Experimental and CCSD(T)-F12 Heats of Formation. *Phys. Chem. Chem. Phys.* **2012,** *14*, 7724-38.
103. Hughes, T. F.; Friesner, R. A., Correcting Systematic Errors in DFT Spin-Splitting Energetics for Transition Metal Complexes. *J. Chem. Theory Comput.* **2011,** *7*, 19-32.
104. Jiang, W.; Deyonker, N. J.; Determan, J. J.; Wilson, A. K., Toward Accurate Theoretical Thermochemistry of First Row Transition Metal Complexes. *J. Phys. Chem. A* **2012,** *116*, 870-885.
105. Bajusz, D.; Racz, A.; Heberger, K., Why Is Tanimoto Index an Appropriate Choice for Fingerprint-Based Similarity Calculations? *J. Cheminf.* **2015,** *7*, 20.
106. O'Boyle, N. M.; Banck, M.; James, C. A.; Morley, C.; Vandermeersch, T.; Hutchison, G. R., Open Babel: An Open Chemical Toolbox. *J. Cheminf.* **2011,** *3*, 33.
107. The Open Babel Package Version 2.3.1. http://openbabel.org (accessed November 5, 2016).